\documentclass[jcp,twocolumn,amsmath,amsfonts,amssymb,showpacs,final]{revtex4-1}
\usepackage{graphicx,graphicx,epsfig}

\usepackage{graphicx}
\usepackage{bm}
\usepackage[all]{xy}

\newcommand{\qed}{\nobreak \ifvmode \relax \else
      \ifdim\lastskip<1.5em \hskip-\lastskip
      \hskip1.5em plus0em minus0.5em \fi \nobreak
      \vrule height0.75em width0.5em depth0.25em\fi}
\newcommand{\beq}{{\begin{equation}}}
\newcommand{\eeq}{{\end{equation}}}

\begin{document}
\title{Quantization and Fractional Quantization of Currents in
  Periodically Driven Stochastic Systems I: Average Currents}

\author{Vladimir~Y.~Chernyak}
\affiliation{Department of Chemistry, Wayne State University, 5101 Cass Avenue, Detroit, MI 48202, Department of Mathematics, Wayne State University, 656 W. Kirby, Detroit, MI 48202, and Theoretical Division, Los Alamos National Laboratory, Los Alamos,
NM 87545 USA}
\author{John~R.~Klein}
\affiliation{Department of Mathematics, Wayne State University, 656 W. Kirby, Detroit, MI 48202,
Center for Nonlinear Studies, Los Alamos National Laboratory, Los Alamos, NM 87545 USA, and New Mexico Consortium, Los Alamos, NM 87545 USA}
\author{Nikolai A.~Sinitsyn}
\affiliation{Theoretical
Division, Los Alamos National Laboratory, Los Alamos, NM 87545 USA and New Mexico Consortium, Los Alamos, NM 87545 USA}


\begin{abstract}
This article studies Markovian stochastic motion of a particle on a graph with finite number of nodes and periodically time-dependent transition rates that satisfy the detailed balance condition at any time. We show that under general conditions, the currents in the system on average become quantized or fractionally quantized for adiabatic driving at sufficiently low temperature. We develop the quantitative theory of  this quantization and interpret it in terms of topological invariants. By implementing the celebrated Kirchhoff theorem we derive a general and explicit formula for the average generated current that plays a role of an efficient tool for treating the current quantization effects.
\end{abstract}

\date{\today}

\maketitle

\section{Introduction}
\label{intro}

This is the first in the series of two articles devoted to the properties of classical mesoscopic
stochastic systems under the influence of adiabatically slow driving. At thermodynamic equilibrium, laws of thermodynamics do not allow such systems to perform a directed motion on average and only random thermal fluctuations can happen. Periodic changes of parameters, however, induce pump and ratchet effects \cite{westerhoff-86} that lead to a directed motion of a system. It is often the case that the average currents in such  systems are
quantized with the quantum numbers
being integers or fractional numbers. This phenomenon was initially discussed for applications to mesoscopic quantum mechanical systems at zero temperature \cite{brouwer-98, makhlin-mirlin-01,moskalets-buttiker-02,pump_berry,kamenev,levitov-02}. More recently, quantization of currents in purely classical mesoscopic systems has also attracted attention.  It has been discussed for a number of applications, including electronic turnstiles \cite{turnstile,buttiker-1},   ratchets  \cite{shi}, molecular motors \cite{astumian-07pnas,astumian-11rev,Leigh-03,sinitsyn-09review,sinitsyn-09jcp,sinitsyn-11jcp}, and heat pumps \cite{hanggi-10}.

Recently it has been  proposed \cite{sinitsyn-09jcp} that this phenomenon occurs generically in stochastic kinetics,
where we have argued that Markovian evolution of a particle in an arbitrary graph leads either to integer or fractional quantization of the average current, generated over long time in the limit of adiabatically slow parameter driving, followed by the low-temperature limit (the latter limit is also equivalent to the condition that the amplitude of perturbation has much larger energy than $k_BT$). In the first (present) manuscript of the series we consider in some detail the quantization of average currents phenomenon, whereas the second manuscript focuses on the full counting statistics of currents in the regime of average current quantization, and relate the latter to specific topological invariants.
%

There are numerous reasons to explore the (fractional) quantization of currents in stochastic systems.Ê Models of periodically driven stochastic systems have applications to chemical kinetics and electronics \cite{westerhoff-86}. 
They have became a major research direction in the field of stochastic thermodynamics, which explores the performance of microscopic engines beyond standard macroscopic thermodynamics framework \cite{sinitsyn-11jpa}.  Quantization of  mesoscopic systems response to periodic driving provides an opportunity for robust control of these systems, e.g. for creating a source of well-controlled current \cite{turnstile}.
In second article of the series, we show that the quantization of currents is related to some fundamental topological properties of the full counting statistics  (FCS) of generated currents. FCS is an active field of research in mesoscopic physics with applications to single molecule experiments and nanoscale electric circuits \cite{cao,makhlin-mirlin-01}. FCS measurements can reveal  information that cannot be obtained by looking at average characteristics. For example, it is known that full counting statistics can be related to  mutual quantum information \cite{FCS-QI} and can show unusual phase transitions in response to periodic driving of parameters \cite{phase-tr-FCS,abanov-10epl,sinitsyn-10jstat}. Exact results for currents and current fluctuations in systems driven periodically by arbitrarily strong and fast driving protocols have been established \cite{sinitsyn-11pre,jarzynski-08prl,sinitsyn-08prl,netocny-10,horowitz-09,sinitsyn-ren-11jstat,jarzynski-11jstat}. Some of the exotic phenomena, such as charge fractionalization, can appear in the counting statistics of the most simple stochastic models \cite{pistolesi,astumian-07pnas,sinitsyn-09jcp}. We shall identify and study a  topologically nontrivial structure in the full counting statistics.
Topological phenomena in physics are a subject of a ongoing interest. Studies of average quantized currents can result in the classification of such new topological phenomena in statistical physics.

We will also show that the master operators that describe the evolution of currents and populations in our models are Hermitian up to a transformation that does not influence the topological properties of their eigenstates. This provides a bridge between  the stochastic phenomena appearing in our work and quantum mechanics.  For example, it implies that all topologically protected numbers that we find in stochastic models, in principle, can be found in some quantum mechanical systems. Classification of quantum mechanical pump effects is currently an important research direction \cite{pump-class} and our work provides a classification for a subclass of such systems.

The main idea that stands behind the mechanism of integer current quantization in the low-temperature adiabatic limit can be described qualitatively in the following way. When the time-dependent rates $k_{i\alpha}$, of a transition from node $i$ via link $\alpha$ of our network are in detailed balance at any time, they can be parameterized [see Eq.~(\ref{rate-detailed-balance})] by the node energies $E_{i}$ and barrier heights $W_{\alpha}$, associated with the network nodes and links, respectively. In the case of adiabatic driving at low temperatures the particle sits at the node $i$ with the lowest value of $E_{i}$, provided the energy is non-degenerate, and therefore no current is generated while the lowest-energy state is non-degenerate. Stated differently the currents can be generated only when the lowest energy $E_{i}$  gets replaced by $E_{j}$ for some other node $j$, which necessarily involves passing through a degeneracy $E_{i}=E_{j}$. Passing through such a degeneracy along a driving protocol is typical, since the latter is represented by a closed curve in the parameter space. The current, responsible for the population flow from node $i$ to node $j$, is generated around the $E_{i}=E_{j}$ degeneracy. However, if the barrier values are non-degenerate during the aforementioned population redistribution, in the low-temperature limit the particle will avoid the higher barriers, so that its motion will be restricted to a particular spanning tree of our network, obtained by withdrawing the links with the highest barrier values, as described in subsection~\ref{minimal-spanning}. Therefore, any path that connects the node $i$ to $j$ is restricted to the tree. Any such a path can bounce many times through links on the tree but contributions to currents from transitions back and forward through the same link cancel. Hence the contributions to current, produced by paths connecting $i$ to $j$ on the tree, are the same. 
We will say that such paths, which are different only by multiple transitions through some links in both directions, belong to the same {\it topological class}. 
Since simultaneous degeneracy of the node energies and barrier heights is non-typical (by the dimensional arguments), for a generic driving protocol the population redistribution involves the paths of the same  topological class. Therefore, when over a period of the driving protocol the particle returns to the starting node, the current over the period is generated by closed paths of the same topological class. Since the generated current depends only on the topological class of the corresponding stochastic path, all relevant stochastic paths, i.e., the ones that provide a substantial contribution to the current can be obtained by one representative. Stated differently, the integral current ${\bm Q}$ (i.e. the flux) that is generated during a period of driving may be viewed as having been produced by a single closed path and is therefore integer valued.

To handle a more involved and technical case of fractional current quantization, which corresponds to a situation with permanent degeneracy of some energies and barriers, we will map the problem of identification of the average instant current ${\bm J}$ (so that the flux ${\bm Q}$ is given by the integral of ${\bm J}(t)$ over the period of driving) onto the problem of finding the currents flowing over the links in an electrical network, given the values of the external currents associated with the nodes; the described problem is known as the Kirchhoff problem \cite{hill-book,graph-book}. This will allow us to apply the Kirchhoff theorem \cite{hill-book,graph-book} that provides an explicit solution of the Kirchhoff problem. 
We will show below that this solution is a powerful and convenient tool in studying the low-temperature adiabatic limit.

The manuscript is organized as follows. In section~\ref{language} we introduce the key objects that play an important role in  both papers of the series. Section~\ref{main-results} contains a preliminary discussion of the main results of the present manuscript for arbitrary connected finite graph. In section~\ref{fractional-Q}, we develop an analogy between the stochastic pump effect and currents in an electrical network. Before we proceed to the derivation of the main results for an arbitrary graph, section~\ref{toy-model} works out in a self-consistent way a simple ``toy-model'' that we use for illustrating the basic concepts and phenomena. Sections~\ref{integer-Q} and \ref{average-Q-rational} provide the explicit derivation of our main results: expressions for the integer and the fractionally quantized currents, respectively. Section~\ref{winding-index} is devoted to topological interpretation of the (fractionally) quantized currents in terms of the winding index around elements of what we call the space of bad parameters. We summarize our findings in section~\ref{conclusion}.

\section{Key Objects and Terminology}
\label{language}

In this section we introduce the key objects appearing in both manuscripts of the series.

\subsection{Markov-chain model for stochastic pumping}
\label{model}

\begin{figure}
\centerline{\includegraphics[width=2.2in]{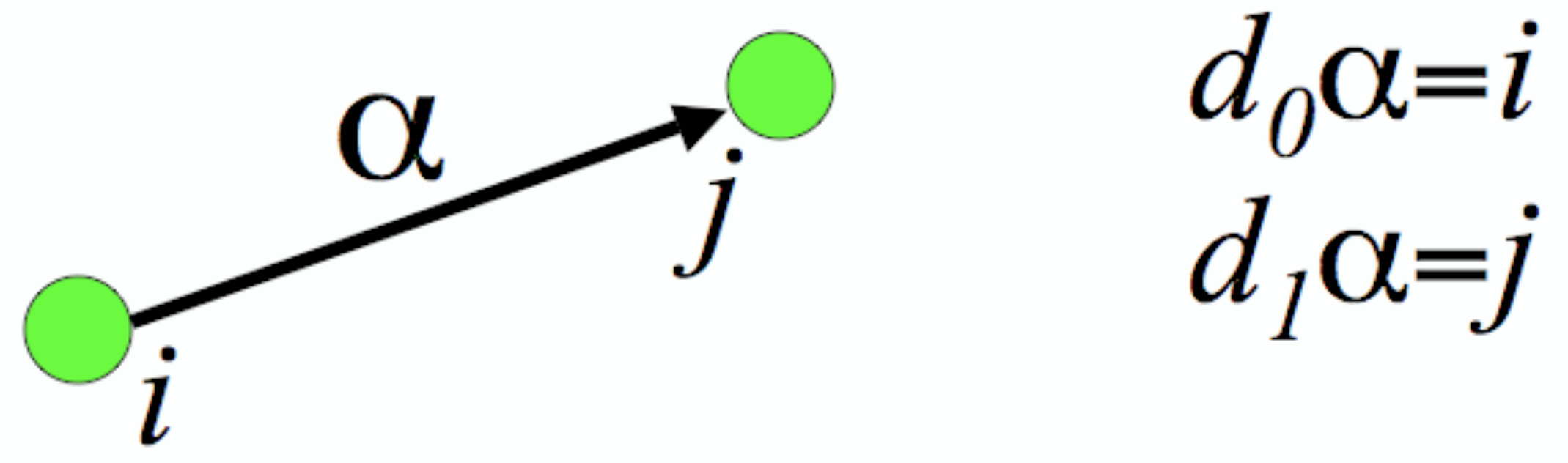}}
  \caption{ Link $\alpha$ between two sites $i$ and $j$. According to our notation,
 $\partial \alpha =\{ i, j \}$,  $d_0 \alpha = \{ i \}$, and  $d_1 \alpha = \{ j \}$.
 }
 \label{link}
 \end{figure}

We start with describing a model of stochastic driving, which
basically follows the instantaneously detailed-balanced Markov-chain
model described in \cite{sinitsyn-08prl} with some slight differences. 
In this instance, the temperature will be treated explicitly. We will also allow more than
one link that connects two nodes, as well as links that connect a node
to itself (self-links).
More formally, our network is described by a non-oriented graph
$X=(X_{0},X_{1})$ with $X_{0}$ and $X_{1}$ being the sets of nodes and
links, respectively. The
non-oriented boundary map, $\partial$,
 associates with a link $\alpha$ its non-oriented boundary
 $\partial\alpha=\{i,j\}$ (or $\partial\alpha=\{j\}$ in the case of a
 self-link), represented by the set of nodes, the link is attached
 to, as we show in Fig.~\ref{link}. It is very convenient to fix some arbitrary
orientation on the graph, i.e., put an arrow (direction) on each
link. This allows oriented boundaries of links to be introduced, which we will represent by standard brackets, i.e. $\partial {\alpha}=(d_{0}\alpha,d_{1}\alpha)$, so that if
$\alpha$ is not a self-link, the arrow on it
goes from $d_{0}\alpha$ to $d_{1}\alpha$; for a self-link
$d_{0}\alpha=d_{1}\alpha$.

A stochastic process is described by a set
$k_{i \alpha}$ of transition rates. The rates can
be (and in the case of stochastic driving actually are)
time-dependent.
\begin{figure}
\centerline{\includegraphics[width=2.2in]{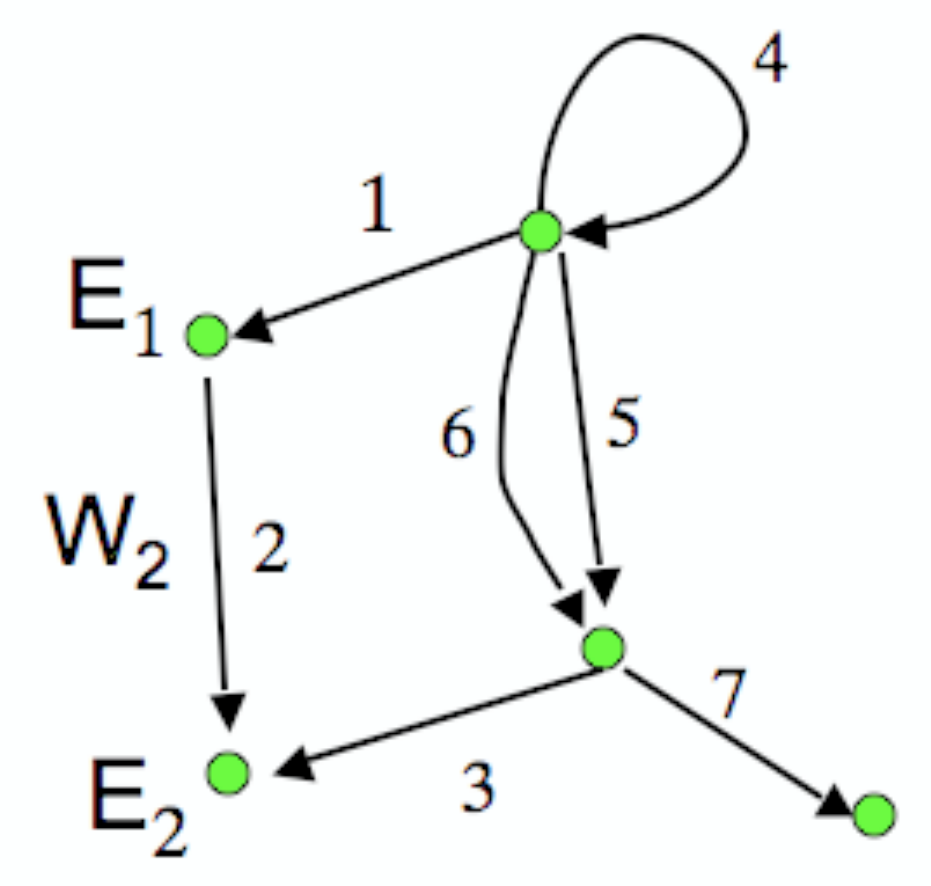}}
  \caption{ A closed graph with 5 nodes and 7 links, representing a Markov chain with five 
 $k_{i\alpha}=k e^{\beta (E_i-W_{\alpha})}$. Nodes are characterized by energies of well 
 Links are characterized by sizes of barriers $W_{\alpha}$, where $\alpha=1,\ldots, 7$.
}
\label{2level-split-1}
\end{figure}
Stochastic dynamics can be described by the master equation for distributions ${\bm\rho}=(\rho_{j}|j\in X_{0})$ that has the form  
\begin{eqnarray}
\label{master-eq} d_{t}{\bm\rho}(t)=\hat{H}(t){\bm\rho}(t), \;\;\; d_{t}\equiv d/dt,
\end{eqnarray} 
where the {\it master operator} $\hat{H}$ is given by
\begin{eqnarray}
\label{master-operator} (\hat{H}{\bm\rho})_{i}=\sum_{j\ne i}
\sum_{\alpha\in X_{1}}^{\partial\alpha=\{i,j\}}k_{j\alpha}\rho_{j}-\bar{k}_{i}\rho_{i},
\end{eqnarray}
with the departure rate
\begin{eqnarray}
\label{rate-departure} \bar{k}_{i}\equiv\sum_{j\ne i}
\sum_{\alpha\in X_{1}}^{\partial\alpha=\{i,j\}}k_{i\alpha}.
\end{eqnarray}
Here the notation,  $\sum_{\alpha\in X_{1}}^{\partial\alpha=\{i,j\}}$
should be understood as the sum over all links having the boundary
represented by nodes $\{i \}$
and $\{j \}$.

\subsection{Detailed balance and Arrhenius parametrization}
We will focus on a situation when at each particular time moment the transition rates are in detailed balance, which means that they can be parameterized as
\begin{eqnarray}
\label{rate-detailed-balance} k_{i\alpha}=ke^{-\beta(W_{\alpha}-E_{i})},
\end{eqnarray}
where $k_{i\alpha}$ is the rate of a jump over the link $\alpha$, starting with node $i$, and $\beta=1/(k_{B}T)$, with $k_{B}$ and $T$ being the Boltzmann constant and temperature, respectively, and $k$ is the constant that insures a proper dimensionality. The choice of this constant does not influence the physics discussed below so, in what follows, we will set $k=1$. The parameters $E_{i}$ and $W_{\alpha}$ have the physical meaning of the potential well energies and the energy positions of the barriers, respectively \cite{jarzynski-08prl,sinitsyn-08prl}. Thus, we will call parameter $E_{i}$ the {\it energy of node i} and we will call $W_{\alpha}$ the {\it barrier height at link $\alpha$}. The master operator depends on time $\hat{H}(t)=\hat{H}({\bm x}(t))$ via its dependence on the set ${\bm x}=({\bm E},{\bm W})$ of parameters.

In the detailed balance case the zero mode of the master operator $\hat{H}$ is usually referred to as the equilibrium distribution, and is represented by the Boltzmann distribution ${\bm\rho}^{B}$, given by
\begin{eqnarray}
\label{Boltzmann-distribution} \rho_{j}^{B}=Z^{-1}e^{-\beta E_{j}}, \;\;\; Z=\sum_{k}e^{-\beta E_{k}}.
\end{eqnarray}
The non-zero eigenmodes ${\bm\xi}$ of $\hat{H}$ that describe relaxation (and are therefore referred to as relaxation modes) satisfy the zero total population condition $\sum_{k}\xi_{k}=0$. Henceforth, we will refer to the subspace of population vectors with zero total population as the space of relaxation modes or,  simply, the relaxation space of $\hat{H}$. Obviously, the relaxation space is an invariant subspace of $\hat{H}$.

We note that, strictly speaking, the expression for the rates should have a form $k_{i\alpha}=e^{-\beta(W_{\alpha}-E_{i})+S_{i}-S_{\alpha}}$, where $S_{i}$ and $S_{\alpha}$ are the entropy terms corresponding to the metastable state, $i$, and transition state, $\alpha$, respectively, however, for the sake of convenience the entropy terms will be absorbed by assuming that $E_{i}$ is the free energy in state $i$ etc. 

In order to simplify our notation, we will use parameters, $\kappa_j$ and $g_{\alpha}$, given by
\begin{equation}
\kappa_j \equiv e^{\beta E_{j}}, \quad g_{\alpha} \equiv e^{-\beta W_{\alpha}}
\label{kap-g}
\end{equation}
so that $k_{j\alpha}=\kappa_j g_{\alpha} $.

\subsection{Driving protocols and generated currents}
\label{protocols-currents}

Let $\widetilde{{\cal M}}_{X}$ be the space of   parameters ${\bm x}=({\bm E},{\bm W})$, associated with a graph $X$.
In this series of two manuscripts we study the currents generated by changing
the system parameters with time so that at each particular time the
system is in detailed balance. Thus, a driving protocol is a smooth map
from time to the parameter space $\widetilde{{\cal M}}_{X}$. We will focus on
periodic driving, which means that a driving protocol is a periodic
function of time, with the values in $\widetilde{{\cal M}}_{X}$ and time
period $\tau_{D}$. Being specifically interested in the adiabatic
limit we introduce the dimensionless time, $\tau=t/\tau_{D}$
(here $D$ is for ``driving''), so that a driving protocol is fully characterized by its time period $\tau_{D}$ and a periodic function ${\bm s}(\tau)$ with period $1$. Obviously, ${\bm s}(\tau)$ can be conveniently interpreted as a map of a circle of unit length to $\widetilde{{\cal M}}_{X}$. The master equation adopts a form
\begin{eqnarray}
\label{master-eq-proper} d_{\tau}{\bm\rho}(\tau)=\tau_{D}\hat{H}({\bm s}(\tau)){\bm\rho}(\tau).
\end{eqnarray}

A stochastic (also known as empirical) current is represented by a
vector ${\bm J}=(J_{\alpha}|\alpha\in X_{1})$. For a stochastic
trajectory ${\bm\eta}$ the generated current is given by ${\bm J}={\bm
  q}/\tau$, 
and $q_{\alpha}$ is the number of times the particle jumped over the link $\alpha$ in the direction of the orientation minus the number of times it jumped over the same link in the opposite direction. 

The boundary map $\partial$ can be extended to an operator that acts
on vectors over the links into vectors over the nodes, with the
physical meaning of being a discrete analogue of the divergence
operator. By acting on the vector of currents, ${\bm J}$, by our definition, it gives
\begin{eqnarray}
\label{define-boundary-operator} (\partial{\bm J})_{i}=
\sum_{j\ne i}\left (
\sum_{\alpha\in X_{1}}^{\partial\alpha=(j,i)}J_{\alpha} - \sum_{\alpha\in X_{1}}^{\partial\alpha=(i,j)}J_{\alpha}\right),
\end{eqnarray}
where the summation, $\sum_{\alpha\in X_{1}}^{\partial\alpha=(i,j)}$, should be understood as the sum over all links of the graph $X$ having node $i$ as the origin and pointing to the node $j$.

We use the operator $\partial$ because it allows us to write various expressions that involve current on a graph in an invariant form, i.e. in the form that does not depend on the choice of orientation of links of a graph. It is also very helpful, for the same purpose, to introduce an operator $\partial^{\dagger}$.
The operator $\partial^{\dagger}$ acts on distributions, resulting in
currents. It has a physical meaning of a discrete analogue of the
gradient operator (with minus sign), and is given by
\begin{eqnarray}
\label{boundary-dagger-explicit} (\partial^{\dagger}{\bm\rho})_{\alpha}=\rho_{d_{0}\alpha}-\rho_{d_{1}\alpha} \;\;\; {\rm for} \; d_{0}\alpha \ne d_{1}\alpha,
\end{eqnarray}
and $(\partial^{\dagger}{\bm\rho})_{\alpha}=0$, when $\alpha$ is a self-link.

It is straightforward to verify that the master operator can be recast in a form that resembles a Fokker-Planck operator for Langevin processes, which will turn out to be useful for our applications:
\begin{eqnarray}
\label{master-operator-2} \hat{H}=\partial e^{-\beta\hat{W}}\partial^{\dagger}e^{\beta\hat{E}},
\end{eqnarray}
where $\hat{W}={\rm diag}(W_{1},\ldots W_{|X_{1}|})$ and
$\hat{E}={\rm diag}(E_{1},\ldots E_{|X_{0}|})$ are diagonal matrices
with values of, respectively, barriers and node energies staying on
the main diagonal.
The stochastic current, generated over a time
period $[0,t_{0}]$ by a driving protocol (not necessarily periodic),
${\bm s}(t)$,
and averaged over the stochastic trajectories and initial distribution
${\bm\rho}{(0)}$, is given by
\begin{eqnarray}
\label{J-average-general} {\bm J}(t)= e^{-\beta\hat{W}(t)}\partial^{\dagger}e^{\beta\hat{E}(t)}{\bm\rho}(t),
\end{eqnarray}
where
\begin{eqnarray}
\label{rho-evolution-general} {\bm\rho}(t)={\rm {\hat{T}}}\exp\left(\int_{0}^{t}dt'\hat{H}({\bm s}(t'))\right){\bm\rho}{(0)}
\end{eqnarray}
is a formal solution of the master equation, and ${\rm {\hat{T}}}$
is the time-ordering operator.
We will focus on the long-time asymptotic behavior of the current
generated via a periodic driving protocol.
The average current, ${\bm Q}$, defined as the average vector of fluxes through the graph links per period of the driving protocol, rather than per unit time is given by
\begin{equation}
\label{currentQ-two} {\bm Q}=N^{-1}\int_0^{N\tau_D} dt {\bm J}(t), \quad N
\rightarrow \infty.
\end{equation}
In the $N\to\infty$ limit, when substituting the general expression for ${\bm J}$ [Eq.~(\ref{J-average-general})] into Eq.~(\ref{currentQ-two}) we can replace ${\bm\rho}(t)$ with the periodic solution of the master equation, which results in the following expression for the average current generated per period of the driving protocol
\begin{eqnarray}
\label{Q-average-general} {\bm Q}({\bm s};\beta,\tau_{D})&=&\tau_{D}\int_{0}^{1}d\tau \nonumber \\ &\times& e^{-\beta\hat{W}(\tau)}\partial^{\dagger}e^{\beta\hat{E}(\tau)}{\bm\rho}(\tau),
\end{eqnarray}
with ${\bm\rho}(\tau)$ being the periodic solution of the master equation [Eq.~(\ref{master-eq-proper})], written in terms of dimensionless time. In the present manuscript, we consider only the average current ${\bm Q}({\bm s};\beta,\tau_{D})$, generated over the driving protocol period, with the focus on the low-temperature adiabatic limit. Its fluctuations are studied in the second manuscript of the series.

\subsection{Good and bad parameters}
\label{good-bad-parmeters}

The quantization of currents, which we discuss below, occurs when the driving protocol belongs to a specific subspace of the parameter space. To define this subspace, we  introduce the {\it space} ${\cal M}_{X}\subset \widetilde{{\cal M}}_{X}$ {\it of good parameters for integer quantization}. It is identified by the condition that for $({\bm E},{\bm W})\in {\cal M}_{X}$ there is no simultaneous degeneracy of the lowest node energies and a degeneracy of two barriers. It is convenient to describe ${\cal M}_{X}$ by introducing a discriminant set, $D_{X}$, of {\it  bad parameters} the protocol should avoid for the integer quantization effect to be in place, so that ${\cal M}_{X}=\widetilde{{\cal M}}_{X}-D_{X}$. More formally, we say that $({\bm E},{\bm W})\in D_{X}$, if $W_{\alpha}=W_{\gamma}$ for some distinct edges $\alpha\ne\gamma$ (barrier degeneracy), and $E_{i}=E_{j}\le E_{k}$ for some distinct nodes $i\ne j$ and any other node $k$ (energy degeneracy). We can naturally represent ${\cal M}_{X}=U_{0}\cup U_{1}$, with $U_{0}$ and $U_{1}$ standing for the sets of parameters $({\bm E},{\bm W})$ with no lowest-node energy, and no barrier height degeneracy, respectively.

Imagine now that we restrict our studies to protocols that keep some of the node energies permanently degenerate and/or keep some of the link barriers permanently degenerate with each other. When studying such a setting with permanent degeneracy among some energies $E_{j}$ and some barriers $W_{\alpha}$ it is convenient to introduce the sets ${\cal X}_{0}$ and ${\cal X}_{1}$ of the subsets ${\cal J}\subset {{X}_{0}}$ and $\mathfrak{A}\subset {X}_{1}$ of the sets of nodes and links, respectively, with permanently identical (degenerate) values of the energies and the barriers. Obviously $X_{0}=\sqcup_{{\cal J}\in {\cal X}_{0}}{\cal J}$, and $X_{1}=\sqcup_{\mathfrak{A}\in {\cal X}_{1}}\mathfrak{A}$. We denote $E_{{\cal J}}$ and $W_{\mathfrak{A}}$ the values of the (degenerate) energies and barriers in ${\cal J}$ and $\mathfrak{A}$, respectively. We can extend the definition of good parameters to this case and introduce the {\it space} ${\cal M}_{X}\subset \widetilde{{\cal M}}_{X}$ {\it of good parameters for rational quantization}, by forbidding simultaneous degeneracy among $E_{{\cal J}}$ and $W_{\mathfrak{A}}$. Similar to the integer quantization case, we have ${\cal M}_{X}=U_{0}\cup U_{1}$.

\subsection{Spanning trees, associated integer currents, and Boltzmann distributions}
\label{spanning-current-Boltzmann}

\begin{figure}
 \centerline{\includegraphics[width=2.8in]{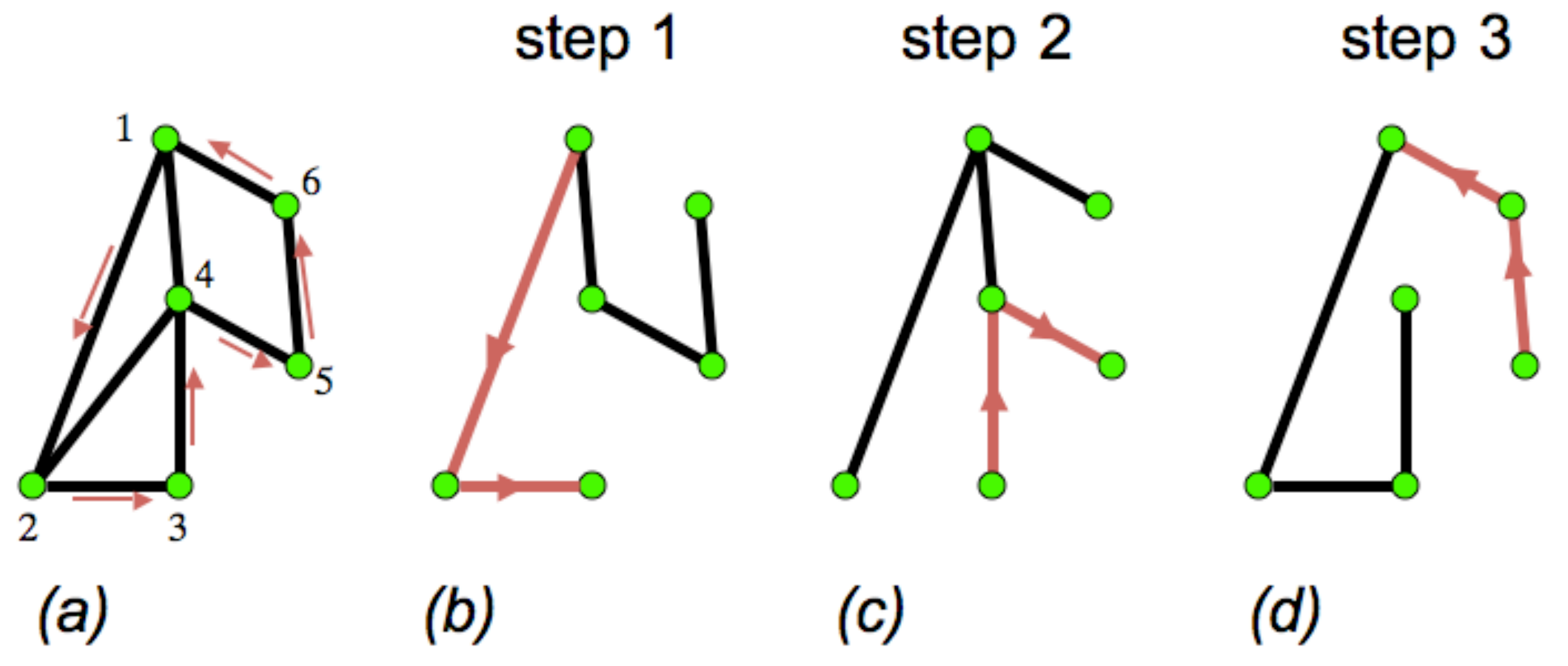}}
  \caption{ (a) Closed path along the contour $1\rightarrow 2 \cdots
    \rightarrow 6$. (b-d) The collection of minimal spanning tree graphs representing  connected components obtained by
 successive elimination of links with highest barriers. Orange color
 indicates the paths on the trees, $l_{ij}(\tilde{X}_{{\bm W}})$, that contribute to the total  integer-valued current.}
 \label{network-steps}
 \end{figure}

A {\it tree} is a connected graph without loops. A {\it spanning tree} of a
graph  is a tree subgraph that contains every vertex of a
graph. Fig.~\ref{network-steps} shows an example of a graph and three
examples of spanning trees of this graph.

For two distinct nodes of $X$ and a spanning tree, $\tilde{X} \in X$, we can identify the unique path, $l_{ij}(\tilde{X})$, that connects $i$ to $j$ through the spanning tree $\tilde{X}$. We also denote by ${\bm Q}_{ij}(\tilde{X})$ the integer-valued current, associated with the described above path. More formally, a component $Q_{ij,\alpha}(\tilde{X})=\pm 1$, when the link $\alpha=\{k,k'\}$ belongs to the path $l_{ij}(\tilde{X})$ with the orienting arrow going $k\to k'$ and $k'\to k$, respectively; and $Q_{ij,\alpha}(\tilde{X})=0$, otherwise.  For example, in Fig.~\ref{network-steps}(b), the orange color shows the path $l_{13}(\tilde{X}_{(b)})$ that generates a unit current through links $(12)$ and $(23)$. More generally, we denote by ${\bm Q}(l)$ the integer-valued current, associated with an oriented path $l$ in our graph. Within the introduced notation we have, in particular ${\bm Q}_{ij}(\tilde{X})={\bm Q}(l_{ij}(\tilde{X}))$.

We further introduce the energy $W(\tilde{X})=\sum_{\alpha\in\tilde{X}_{1}}W_{\alpha}$, associated with a spanning tree $\tilde{X}$, and the Boltzmann distribution ${\bm\varrho}^{B}$ in the space of spanning trees
\begin{eqnarray}
\label{Boltzmann-distr-trees} \varrho_{\tilde{X}}^{B}={\cal Z}^{-1}e^{-\beta W(\tilde{X})}, \;\;\; {\cal Z}=\sum_{\tilde{X}}e^{-\beta W(\tilde{X})}.
\end{eqnarray}

The integer-valued currents $\bm{Q}_{ij}(\tilde{X})$ and the Boltzmann distributions ${\bm\varrho}^{B}$ are important ingredients of the explicit expressions for the average currents, derived in this paper.

\subsection{Conserved currents, weighted scalar products, and non-circulating currents}
\label{conserved-weighted}

In this article we will frequently encounter {\it conserved currents} that are formally defined by the condition $\partial {\bm J}=0$, meaning that the sum of incoming minus the sum of outgoing currents at any node is zero. The pump current per cycle of the driving protocol (\ref{Q-average-general}) is an example of a conserved current. Indeed, if it were not true, then by repeating the cyclic protocol we would be able to increase the population of some nodes indefinitely. To avoid this inconsistency we have to assume that
\begin{equation}
\partial {\bm Q}({\bm s};\beta,\tau_{D}) = 0,
\label{q-cons}
\end{equation}
i.e. the sum of going through any node pump currents is zero. This condition restricts the number of possible independent pump currents to the number of independent loops of the graph. This number is also equal to the number of links that can be removed from the graph without breaking this graph into disjoint components.

Another example of a class of conserved currents, important for our applications, is represented by the currents ${\bm Q}(l)$, associated with paths $l$ in the case when $l$ is closed, i.e., it is a loop. It is evident that ${\bm Q}(l)$ is a conserved current  when $l$ is a loop. We can say even more: the vector space of conserved currents is generated by ${\bm Q}(l)$. A natural way to construct a basis set in the vector space of conserved currents starts with choosing a spanning tree $\tilde{X}\subset X$ of our graphs. Each link $\alpha\in X_{1}\setminus \tilde{X}_{1}$ that does not belong to the tree gives rise to a loop $l_{\alpha}(\tilde{X})$ obtained by concatenating the link $\alpha$ with the only non-self-intersecting path that connects its ends via the tree $\tilde{X}$. The set $\{l_{\alpha}(\tilde{X})|\alpha\in X_{1}\setminus \tilde{X}_{1}\}$ forms a basis set in the space of conserved currents.

Conserved currents naturally do not re-distribute the population between the nodes. We will show that, when populations are re-distributed, the (non-conserved) currents responsible for this process should minimize the circulating components (i.e., represented by conserved currents), which can be formulated as orthogonality to all conserved currents with respect to a {\it weighted scalar product} with conserved currents, defined by
\begin{equation} ({\bm J},{\bm J}')_{g} \equiv  \sum_{\alpha \in X_{1}} (g_{\alpha})^{-1}J_{\alpha}J'_{\alpha}=\sum_{\alpha \in X_{1}}e^{\beta W_{\alpha}} J_{\alpha}J'_{\alpha}.
\label{scal-weigh-def}
\end{equation}
Hereafter, we refer to the currents orthogonal to the conserved counterparts, in the sense of weighted scalar product, given by Eq.~(\ref{scal-weigh-def}) as {\it non-circulating} currents. Since, as stated before, conserved currents are generated by the currents, generated by the loops, the non-circulating (orthogonality) condition for a current ${\bm J}$ reads
\begin{equation} ({\bm Q}(l),{\bm J})_{g} \equiv  \sum_{\alpha \in X_{1}} Q_{\alpha}(l)(g_{\alpha})^{-1}J_{\alpha}=0,
\label{scal-weigh-def-2}
\end{equation}
for all closed paths (loops) in our graph. If we interpret $g_{\alpha}$ and $g_{\alpha}^{-1}$ as the conductivity and resistance, respectively, of link $\alpha$, followed by naturally interpreting $(g_{\alpha})^{-1}J_{\alpha}$ as the voltage on link $\alpha$, then Eq.~(\ref{scal-weigh-def-2}) claims that the sum of voltages of the links over closed path $l$ is zero. Note that the factors $Q_{\alpha}(l)=0,\pm 1$ restrict the summation in Eq.~(\ref{scal-weigh-def-2}) to the links that belong to the path, as well as ensure that the voltages (potential differences) are always taken in the direction of the path, as they should be.

\subsection{Vector-potential and geometric properties of pump currents}
\label{Q-via-A}

It has been demonstrated previously \cite{jarzynski-08prl} that in the case when Arrhenius parametrization is fulfilled at all times, the pump current is totally determined by the geometry of a contour in the space of population vectors, i.e.,
\begin{equation}
{\bm Q}({\bm s};\beta,\tau_{D})=\int_{\bm s} \sum_{k} {\bm A}^{k} d\rho_{k},
\label{q-geom2}
\end{equation}
where the vector-potential, ${\bm A}$, is a vector in the space of currents (i.e., the vector space spanned on the set $X_{1}$ of the graph links) and a linear functional in the space of population vectors with zero total population. Stated equivalently, ${\bm A}$ is a linear operator that maps the population vectors with zero total population to the currents. We further note that, in the adiabatic limit, the formula (\ref{q-geom2}) is almost obvious. Indeed, for a time-independent case the system is at equilibrium, represented by the Boltzmann distribution ${\bm\rho}^{B}$ with zero average currents. When parameters are adiabatically time-dependent,  the instantaneous average current, to the lowest order in time-derivatives,  must be a linear combination of first time-derivative of the equilibrium node populations, i.e. ${\bm J}=\sum_{k} {\bm A}^{k}\dot{\rho}_{k}^{B}$. Integrating over time of the protocol we find that in the adiabatic limit the total current is totally determined by the protocol geometry and does not depend on time explicitly, that is,
\begin{equation}
{\bm Q}({\bm s};\beta)=\lim_{\tau_{D}\to\infty}{\bm Q}({\bm s};\beta,\tau_{D})=\int_{\bm s} \sum_{k} {\bm A}^{k}d\rho_{k}^{B}.
\label{q-geom}
\end{equation}

At this point we would like to note that, while ${\bm A}$ is described as a linear operator in an unambiguous way, the description of its components ${\bm A}^{k}$ needs some clarification, due to restricted nature of the operator domain. Denoting by $|j\rangle$ the population vectors with $\rho_{k}=\delta_{kj}$, and introducing a reference node $a\in X_{0}$, we can see that the set of vectors $|j\rangle-|a\rangle$, with $j\in X_{0}$ and $j\ne a$ form a basis set in the space of population vectors with zero total population. We further naturally define ${\bm A}_{a}^{k}={\bm A}(|k\rangle-|a\rangle)$. Obviously, the set $({\bm A}_{a}^{k}|k\in X_{0})$ of components depends parametrically on the choice of the reference node $a$, however, $\sum_{k}{\bm A}_{a}^{k}d\rho_{k}$ is $a$-independent, since $\sum_{k}d\rho_{k}=0$ for normalized distributions. Therefore, the subscript $a$ can be dropped in the expressions like Eqs.~(\ref{q-geom2}) and (\ref{q-geom}). Also the notation ${\bm A}^k-{\bm A}^{k'}$ is free of the above ambiguity, since ${\bm A}_{a}^k-{\bm A}_{a}^{k'}={\bm A}_{k'}^{k}$ are $a$-independent.
%


\section{Main Results}
\label{main-results}

In this first manuscript of the series we study the average current ${\bm Q}({\bm s};\beta,\tau_{D})$, generated during the period of a driving protocol in networks, represented by arbitrary graphs, with the main focus on its adiabatic followed by the low-temperature limit  $\bar{{\bm Q}}({\bm s})= \lim_{\beta\to\infty}\lim_{\tau_{D}\to\infty}{\bm Q}({\bm s};\beta,\tau_{D})$.
Recently we have proposed the so-called  {\it Pumping-Quantization Theorem} (PQT) \cite{sinitsyn-09jcp} that describes the average currents generated per driving period, and can be formulated as follows:

(i) \underline{Integer PQT}: For an arbitrary graph and under the condition that the periodic driving protocol always stays inside the subspace of good parameters for integer quantization (which was defined in subsection~\ref{good-bad-parmeters}), the average current per cycle must be quantized in the adiabatic limit, which is followed by the low temperature limit.

(ii) \underline{Fractional PQT}: If we allow permanent degeneracies of some barriers and/or energies and assume that the driving protocol belongs to the space of good parameters for rational quantization, while keeping all other conditions of (i) fulfilled, the current per cycle of the driving protocol is
a rational (fractional) number.

The main result of this first manuscript of our series is the  {\it averaging formula} (AF) that quantifies the phenomenon of (fractional) quantization of currents. The AF states that  the average current,  $\bar{{\bm Q}}({\bm s})$, per cycle, ${\bm s}$, of the driving protocol can be written explicitly in the following form:
\begin{eqnarray}
\label{Q-averaging55} \bar{{\bm Q}}({\bm s})=
\frac{1}{n_{X}}\sum_{({\bm k},\tilde{{\bm X}})}\bar{{\bm Q}}({\bm s};{\bm k},\tilde{{\bm X}}),
\end{eqnarray}
that expresses $\bar{{\bm Q}}({\bm s})$ as a normalized sum of the integer valued conserved currents, $\bar{{\bm Q}}({\bm s};{\bm k},\tilde{{\bm X}})$, associated with $({\bm k},\tilde{{\bm X}})$, which represents all possible ways to split specific parameter degeneracies that are encountered along the driving protocol, referred to as {\it global degeneracy resolutions}, with the normalized integer denominator $n_{X}$ that represents the number of legitimate global degeneracy resolutions. The explicit construction for $\bar{{\bm Q}}({\bm s};{\bm k},\tilde{{\bm X}})$ and definitions of $n_{X}$ and $({\bm k},\tilde{{\bm X}})$ are provided in section~\ref{average-Q-rational}. At this point we would like to emphasize that the AF for the average current not only delivers a proof of the fractional PQT, but also provides an explicit construction that allows the average current to be computed in the low-temperature adiabatic regime. Note that integer quantization can be viewed as a particular case of rational quantization when there is no degeneracy, which implies $n_{X}=1$. We will first consider in section~\ref{integer-Q} this special case of integer quantization in some detail, while the more general case of rational quantization is postponed to section~\ref{average-Q-rational}. There are two main reasons for that. First, the integer-valued currents $\bar{{\bm Q}}({\bm s};{\bm k},\tilde{{\bm X}})$ in the r.h.s. of the AF [Eq.~(\ref{Q-averaging55})] can be viewed as integer-quantized currents that correspond to legitimate ways, labeled by $({\bm k},{{\bm X}})$, to resolve the degeneracy, so that understanding the nature of integer quantization provides an additional insight into the rational quantization phenomenon. Second, at the present point the nature of integer quantization is understood on a deeper level of the current probability distribution, or more specifically the associated with it large deviation function ${\cal S}({\bm Q})$, as presented with some detail in the second manuscript of the series.


On the way to the derivation of the AF, we will obtain several results that are of independent interest in the theory of the pump effect. Thus the derivation of the AF [Eq.~(\ref{Q-averaging55})] includes the following steps:

1) We show that components of  the vector-potential, ${\bm A}$ in Eq.~(\ref{q-geom2}) can be
identified with the current distributions on a graph of an electrical network, in which the links
represent the resistances $r_{\alpha}$, given by
\begin{equation}
r_{\alpha}=g_{\alpha}^{-1}=e^{\beta W_{\alpha}},
\label{RRR}
\end{equation}
so that the component, ${\bm A}_{a}^{k}$, corresponds to current vector in our network that would appear if a unit current enters the network and leaves it at nodes $a$ and $k$, respectively.

2) We show [Eqs.~(\ref{A-explicit}) and (\ref{A-Kirchhoff-general})] that  the  identification of ${\bm A}$ with a current in a resistance network allows us to apply the Kirchhoff's theorem and determine the expressions for ${\bm A}_{a}^{k}$ and ${\bm A}^{k'}-{\bm A}^{k}$ explicitly in terms of the Boltzmann weights of all spanning trees and currents that correspond to paths that connect nodes $k$ and $k'$ on these trees. More specifically, we introduce an energy function on the set of spanning trees that associates with a spanning tree $\tilde{X}$ the energy $W(\tilde{X})$ given by the sum of the barrier heights of the links that belong to $\tilde{X}$, as well as the corresponding Boltzmann distribution ${\bm\varrho}^{B}$, given by Eq.~(\ref{Boltzmann-distr-trees}).

3) The above identification leads to a general and explicit expression [Eq.~(\ref{q-geom2-explicit})] for the average current ${\bm Q}({\bm s};\beta,\tau_{D})$ in terms of the tensor $\bm{Q}^{k\tilde{X}}$ (with values in the integer-valued currents, and superscripts $k$ and $\tilde{X}$ running over the graph nodes and spanning trees, respectively) that depends on the graph structure only, and a distribution $\bm{\gamma}$ in the combined space of nodes and spanning trees. Explicit integral expressions [Eqs.~(\ref{define-gamma-distr}) and (\ref{gamma-distr-alterntive})] for $\bm{\gamma}$ involve only two loops (closed paths) in the spaces of node and spanning tree distributions: these are the periodic solution ${\bm\rho}(\tau)$ of the master equation and the time dependent tree Boltzmann distribution ${\bm\varrho}^{B}(\tau)$ that depends on time via periodic dependence ${\bm W}(\tau)$ of the barrier data, according to the driving protocol.

4) The adiabatic $\tau_{D}\to \infty$ limit is obtained in a straightforward way by replacing ${\bm\rho}(\tau)$ in the expressions for ${\bm\gamma}$ [Eqs.~(\ref{define-gamma-distr}) and (\ref{gamma-distr-alterntive})] with its adiabatic limit, represented by ${\bm\rho}^{B}(\tau)$. The obtained expressions [Eqs.~(\ref{gamma-distr-adiabatic})] show symmetry between the sets of nodes and spanning trees, which might be indicative of some kind of duality in the adiabatic limit, the latter being not yet understood.

After establishing the AF, we develop a {\it winding index} interpretation of the average current. This allows the space ${\cal M}_{X}^{r}$ of {\it robust parameters}, or simply the {\it robust space} to be identified, which constitutes the broadest subspace that provides robust behavior of the pumped currents. This means that, for a periodic driving protocol ${\bm s}$ that stays at all times within the subspace of robust parameters, the limit $\lim_{\beta\to\infty}\lim_{\tau_{D}\to\infty} \bar{{\bm Q}}({\bm s})$ does exist, and the generated current does not change upon any deformation of the driving protocol ${\bm s}$ as long as ${\bm s}$ stays within the robust subspace. In particular, this means that the generated current $\bar{{\bm Q}}({\bm s})$ depends only on the topological class of the driving protocol, considered as a $1$-dimensional cycle in the space of  robust parameters. It is convenient to represent the subspaces ${\cal M}_{X}^{r}=\widetilde{{\cal M}}_{X}-D_{X}^{r}$ as a result of withdrawing the {\it discriminant subset}, $D_{X}^{r}$, of non-robust parameters. The discriminant set has codimension $2$, so it is possible to describe the current in terms of how many times the driving protocol winds around the components of the discriminant set $D_{X}^{r}$. We explore the structure of $D_{X}^{r}$ and show that it consists of a bunch of (codimension $2$) cells, glued together via their borders, and is closely related to the structure of the discriminant $D_{X}$ of bad parameters that was introduced in subsection~\ref{good-bad-parmeters}. The only difference is that certain simultaneous degeneracies do not contribute to the currents and, therefore the corresponding cells can be withdrawn from the discriminant set, resulting in $D_{X}^{r}\subset D_{X}$.

Since the discriminant subspace $D_{X}$, and hence $D_{X}^{r}$, has codimension $2$, robust driving protocols, i.e., the ones that stay within the robust parameter subspace, are typical. This means that, if one creates a protocol with randomly chosen parameters, such protocol will belong to the robust space.

\section{Explicit Expressions for Average Currents and Kirchhoff Theorem}
\label{fractional-Q}

\subsection{Vector potential and Kirchhoff problem}
\label{Kirchhoff-laws}

In this subsection we derive the expression for the average pumped current, given by Eq.~(\ref{q-geom2}) by (a) expressing the current ${\bm J}$, defined by Eq.~(\ref{J-average-general}) as a solution of a linear problem, (b) demonstrate that the latter has a unique solution, which allows us to represent the current in a form ${\bm J}=\sum_{k}{\bm A}^{k}\dot{\rho}_{k}$, (c) We will further demonstrate the equivalence between the aforementioned problem and the Kirchhoff problem of finding the current distribution in an electrical network with the given distribution of the (incoming and outgoing) external currents. This analogy will allow  us to apply the Kirchhoff theorem that provides an explicit form for the solution of the Kirchhoff problem, and represents the linear operator ${\bm A}$ as a weighted sum over all spanning trees of $X$. This will be presented in subsection~\ref{Kirchhoff-theorem}.

The linear problem, discussed above is formulated in terms of two properties of the generated current ${\bm J}$:

(i) the continuity equation $\partial {\bm J}=\dot{{\bm\rho}}$, 

(ii) ${\bm J}$ is non-circulating, i.e., it is orthogonal to all conserved currents with respect to the weighted scalar product, i.e., $({\bm J},{\bm J}_{0})_{g}=0$ for any ${\bm J}_{0}$ with $\partial{\bm J}_{0}=0$.

To establish property (i) we apply the divergence operator to the definition of ${\bm J}$, given by Eq.~(\ref{J-average-general}), followed by making use of the representation of the master operator, given by Eq.~(\ref{master-operator-2}), which results in $\partial{\bm J}=\hat{H}{\bm\rho}=\dot{{\bm\rho}}$. Property (ii) is verified by the following sequence of equalities based on the expression for ${\bm J}$ [Eq.~(\ref{J-average-general})] and definition of a conjugate operator $({\bm J},{\bm J}_{0})_{g}= e^{\beta\hat{W}}{\bm J}\cdot{\bm J}_{0} =\partial^{\dagger} e^{\beta\hat{E}}{\bm\rho}\cdot {\bm J}_{0}=e^{\beta\hat{E}}{\bm\rho}\cdot \partial{\bm J}_{0}=0$.

Provided the linear problem under consideration has a unique solution, it can be represented in a form ${\bm J}={\bm A}{\bm\rho}$, where ${\bm A}$ is a linear operator that acts from the vector space of relaxation modes to the vector space of currents. This immediately establishes the expression for $\bar{{\bm Q}}({\bm s})$, given by Eq.~(\ref{q-geom2}), as well as fully describes the linear operator ${\bm A}$ as providing the solution of the linear problem under discussion.

The uniqueness of the solution of our linear problem is almost obvious. Let ${\bm J}$ and ${\bm J}'$ be its solutions. Then $({\bm J}-{\bm J}')$ is a conserved current orthogonal to all conserved currents, and therefore is equal to zero, i.e., ${\bm J}={\bm J}'$. The existence of a solution follows from the facts that the boundary (discrete divergence) operator $\partial$ maps the vector space of currents onto the relaxation space of $\hat{H}$, and the kernel of $\partial$ coincides with the subspace of conserved currents. Therefore, since the space of non-circulating currents is the orthogonal complement to the space of conserved currents, the operator $\partial$ provides an isomorphism between the space of of non-circulating currents and relaxed populations. Therefore ${\bm A}$ can be chosen as the inverse of the above isomorphism.

To establish equivalence between our linear problem and the Kirchhoff problem, we recast the conditions (i) and (ii) in terms of the components ${\bm A}_{a}^{k}$, i.e., setting $\dot{\rho}_{j}=\delta_{j}^{k}-\delta_{j}^{a}$, so that condition (i) reads
\begin{eqnarray}
\label{A-condition-1} (\partial{\bm A}_{a}^{k})_{j}=\delta_{j}^{k}-\delta_{j}^{a}.
\end{eqnarray}
We further recast condition (ii), i.e., the orthogonality condition using conserved currents, associated with loops $l$ of our graph [Eq.~(\ref{scal-weigh-def-2})]
\begin{eqnarray}
\label{A-condition-2} \sum_{\alpha \in X_{1}}Q_{\alpha}(l)e^{\beta W_{\alpha}}A_{a\alpha}^{k}=0.
\end{eqnarray}

We now interpret $A_{a\alpha}^{k}$ as electrical current that goes over link $\alpha$ of an electrical circuit, represented by our graph, provided we have an incoming and outgoing currents both of value $+1$ at nodes $a$ and $k$, respectively. According to the discussion that followed Eq.~(\ref{scal-weigh-def-2}), we interpret $r_{\alpha}=e^{\beta W_{\alpha}}$ as the resistance of link $\alpha$, which makes $e^{\beta W_{\alpha}}A_{a\alpha}^{k}$ to be the voltage on link $\alpha$. Within the presented interpretation, Eq.~(\ref{A-condition-1}) represents first Kirchhoff law: At any node in an electrical circuit, the sum of currents flowing into that node is equal to the sum of currents flowing out of that node; whereas Eq.~(\ref{A-condition-2}) reflects second Kirchhoff law: The directed sum of the electrical potential differences (voltages) around any closed circuit is zero. Therefore, ${\bm A}_{a}^{k}$ is given by the solution of the Kirchhoff problem: Find the stationary currents that flow over the links of an electrical network, represented by a graph $X$, provided we have an incoming and outgoing external currents both of value $+1$ at nodes $a$ and $k$, respectively.

\subsection{Kirchhoff theorem and explicit expressions for average currents}
\label{Kirchhoff-theorem}

The solution of the Kirchhoff problem is provided by the celebrated Kirchhoff theorem \cite{hill-book,graph-book}. The Kirchhoff theorem provides an explicit expression for the current, in our case denoted ${\bm A}_{a}^{k}$, in a form of 
a weighted sum of integer currents ${\bm Q}_{ak}(\tilde{X})$ (defined in subsection~\ref{spanning-current-Boltzmann}) over all possible spanning trees $\tilde{X}\subset X$ of our graph \cite{hill-book,graph-book}. According to Kirchhoff's theorem, the relative (non-normalized) weight of a spanning tree $\tilde{X}$ is given by the product of conductances (inverse resistances) of its links, the normalization factor ensures the (normalized) weights to sum up to one. Given the above interpretation of $e^{\beta W_{\alpha}}$ as the resistance of link $\alpha$, the definition of the tree energy $W(\tilde{X})$, and the tree Boltzmann distribution $\bm{\varrho}^{B}$ [Eq.~(\ref{Boltzmann-distr-trees})] Kirchhoff's theorem leads to an explicit expression for the vector potential ${\bm A}$
\begin{eqnarray}
\label{A-explicit} {\bm A}_{a}^{k}= \sum_{\tilde{X}}\varrho_{\tilde{X}}^{B}{\bm Q}_{ak} (\tilde{X}),
\end{eqnarray}
which also implies
\begin{eqnarray}
\label{A-Kirchhoff-general} {\bm A}^{k}-{\bm A}^{k'}=\sum_{\tilde{X}}\varrho_{\tilde{X}}^{B}({\bm W},\beta){\bm Q}_{k'k}(\tilde{X}).
\end{eqnarray}
The explicit expression [Eq.~(\ref{A-Kirchhoff-general})] is very convenient for obtaining the low-temperature limit $\beta\to\infty$, which will be explored in the manuscript. We wish to   reiterate that in Eqs.~(\ref{A-explicit}) and (\ref{A-Kirchhoff-general}) summation runs over all spanning trees $\tilde{X}\subset X$ of our graph $X$.

An explicit expression for the average current is obtained upon substitution of Eq.~(\ref{A-explicit}) into Eq.~(\ref{q-geom2}). To emphasize some symmetry between the nodes and spanning trees we also introduce a notation ${\bm Q}_{a}^{k\tilde{X}}={\bm Q}_{ak}(\tilde{X})$, with omitting $a$, when the expressions are independent of the reference node choice. This results in
\begin{eqnarray}
\label{q-geom2-explicit} {\bm Q}({\bm s};\beta,\tau_{D})=\sum_{k,\tilde{X}}{\bm Q}^{k\tilde{X}}\gamma_{k\tilde{X}}({\bm s};\beta,\tau_{D}),
\end{eqnarray}
with
\begin{eqnarray}
\label{define-gamma-distr} \gamma_{k\tilde{X}}({\bm s};\beta,\tau_{D})\equiv \int_{\bm s}\varrho_{\tilde{X}}^{B}d\rho_{k}.
\end{eqnarray}
We can think about ${\bm \gamma}$ as a distribution in the combined space of nodes and spanning trees. The distribution ${\bm\gamma}$ satisfies the following properties:
\begin{eqnarray}
\label{gamma-distr-properties} \sum_{k}\gamma_{k\tilde{X}}=0, \;\;\; \sum_{\tilde{X}}\gamma_{k\tilde{X}}=0.
\end{eqnarray}
The first property allows the reference node to be omitted in the r.h.s. of Eq.~(\ref{q-geom2-explicit}), the second property provides the conserved nature of the average current $\bar{{\bm Q}}({\bm s})$.

Integration by parts and the Stokes theorem provide two alternative expressions for ${\bm\gamma}$
\begin{eqnarray}
\label{gamma-distr-alterntive} \gamma_{k\tilde{X}}({\bm s};\beta,\tau_{D})=-\int_{\bm s}\rho_{k}d\varrho_{\tilde{X}}^{B}=\int_{{\cal A}}d\varrho_{\tilde{X}}^{B}\wedge d\rho_{k},
\end{eqnarray}
where ${\cal A}$ is surface spanned on ${\bm s}$ in the cartesian product of the space of normalized node distributions and the space of normalized tree distributions. Since both distribution spaces are contractible, such a surface always exists.

The derived expressions for the average current are very convenient for deriving its adiabatic limit ${\bm Q}({\bm s};\tau_{D})$, which is achieved by simply replacing ${\bm\gamma}({\bm s};\beta,\tau_{D})$ in Eq.~(\ref{q-geom2-explicit}) by its adiabatic limit ${\bm\gamma}({\bm s},\beta)$, which boils down to replacing the distributions ${\bm\rho}$ in the expressions, given by Eqs.~(\ref{define-gamma-distr}) and (\ref{gamma-distr-alterntive}) with the Boltzmann distributions ${\bm\rho}^{B}$, which results in
\begin{eqnarray}
\label{q-explicit-adiabatic} {\bm Q}({\bm s};\beta)=\sum_{k,\tilde{X}}{\bm Q}^{k\tilde{X}}\gamma_{k\tilde{X}}({\bm s};\beta),
\end{eqnarray}
with
\begin{eqnarray}
\label{gamma-distr-adiabatic} \gamma_{k\tilde{X}}({\bm s};\beta)&=&\int_{\bm s}\varrho_{\tilde{X}}^{B}d\rho_{k}^{B} \nonumber \\ &=&-\int_{\bm s}\rho_{k}^{B}d\varrho_{\tilde{X}}^{B}=\int_{{\cal A}}d\varrho_{\tilde{X}}^{B}\wedge d\rho_{k}^{B}.
\end{eqnarray}
We note that the adiabatic expression for the distribution ${\bm\gamma}$ show symmetry between the sets of nodes and spanning trees of the graph. This might be a signature of certain duality, with the nodes of the dual graph being represented by the spanning trees of the original graph $X$. At this point it is unclear to us whether such duality really takes place, and in particular how to define the links of the dual graph that connect the spanning trees.


\section{Toy model}
\label{toy-model}

\begin{figure}
\centerline{\includegraphics[width=2.0in]{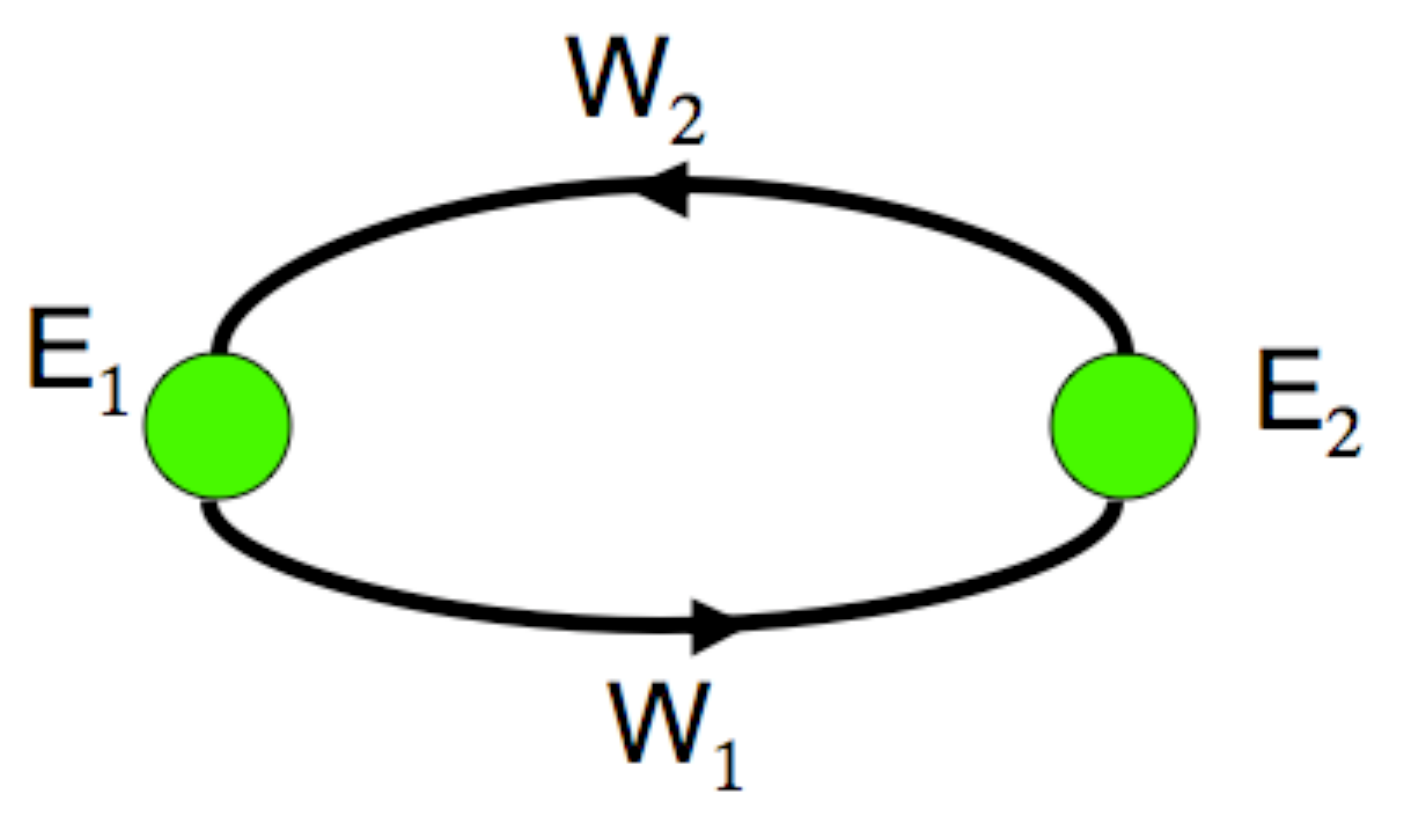}}
  \caption{The ``2-nodes-2-links'' model of stochastic transitions between two states along two possible paths. Different paths are
 characterized by the values of the corresponding potential barriers $W_1$ and $W_2$. The two  states are characterized by
 the values of their well depths, $E_1$ and $E_2$. Periodic
 modulation of these parameters leads to a nonzero, on average, circulating current in the preferred (clockwise or counterclockwise) direction. }
\label{2level-1}
 \end{figure}

As a simple example of a graph of the type shown in Fig.~\ref{2level-split-1}, we consider a model illustrated in Fig.~\ref{2level-1}: a particle can randomly jump between two sites along two different paths. Each path can be traversed in both directions and arrows on the links indicate the directions of particle
transitions that contribute to the currents with the plus sign. This is the minimal model that shows integer current quantization and that we will work out in this section in detail. A reader, who is interested to see further specific examples that include fractional quantization, might want to look at our previous publication \cite{sinitsyn-09jcp}.

\subsection{Markov-chain on a  2-nodes-2-links graph}
\label{model2}

Populations on the nodes in our toy model are described by the population (probability) vector, ${\bm \rho}=(\rho_1,\rho_2)$, where $\rho_1$ and $\rho_2$ are probabilities for a particle to be in the first
and the second nodes, respectively. They evolve according to the master equation [a particular case of Eq.~(\ref{master-eq})],
\begin{equation}
\frac{d}{dt}
\left( \begin{array}{c}
\rho_1 \\
\rho_2
\end{array} \right) = \hat{H} \left( \begin{array}{c}
\rho_1 \\
\rho_2
\end{array} \right).
\label{partial}
\end{equation}
with $\hat{H}$, being the master operator, according to its general definition [Eq.~(\ref{master-operator})].
In terms of the parameters $g_{\alpha}$, with $\alpha =1,2$, and
$\kappa_j$, with $j=1,2$, the master operator $\hat{H}$ reads
\begin{equation}
\hat{H}=
\left( \begin{array}{cc}
-\kappa_1(g_1+g_2) & \kappa_2(g_1+g_2) \\
\kappa_1(g_1+g_2) & -\kappa_2(g_1+g_2)
\end{array} \right).
\label{master-simple}
\end{equation}

\subsection{Driving protocols and generated currents}
\label{toy-protocols-currents}

For the model in Fig.~\ref{2level-1}, the $4$-dimensional parameter space $\widetilde{{\cal M}}$ describes the sets ${\bm x}=(E_1,E_2,W_1,W_2 )$ of controlled parameters. A driving protocol ${\bm s}$ is naturally a cycle in $\widetilde{{\cal M}}$. For example, ${\bm s}$ can be a cycle defined by $E_2=W_2=0$, $E_1(\tau)=E_0\cos(2\pi \tau)$, $W_1(\tau)=W_0\sin(2\pi \tau)$, where $\tau$ is the dimensionless
time given by $\tau =t/\tau_D$; with $\tau_D$ and $t$ being the period of driving and actual time, respectively. The driving protocol is, therefore, completely determined by its duration $\tau_{D}$ and ${\bm s}(\tau)$ for $\tau\in [0,1)$.

In addition to node populations, $\rho_1$ and $\rho_2$, we consider the vector of currents ${\bm J}=(J_1,J_2)$, with $J_1$ and $J_2$ being the random variables that describe how many times per unit time the particle jumps through the corresponding graph links, as shown in Fig.~\ref{2level-1}, counting counterclockwise and clockwise transitions with a plus and minus sign, respectively. The boundary operator $\partial$, which returns the sum of currents outgoing
from each node when it is applied to the vector of currents, takes the form
\begin{equation}
\partial
\left( \begin{array}{c}
J_1 \\
J_2
\end{array} \right) = \left( \begin{array}{c}
J_2-J_1 \\
J_1-J_2
\end{array} \right), \quad \partial = \left(
\begin{array}{rr}
-1 & 1\\
1 & -1
\end{array} \right),
\label{master-two}
\end{equation}
in the full agreement with its general case definition [Eq.~(\ref{define-boundary-operator})].

Its conjugate operator $\partial^{\dagger}$ returns the difference of node populations between the origin and the target nodes of each link, namely,
\begin{equation}
\partial^{\dagger}
\left( \begin{array}{c}
\rho_1 \\
\rho_2
\end{array} \right) = \left( \begin{array}{c}
\rho_1-\rho_2 \\
\rho_2-\rho_1
\end{array} \right), \quad \partial^{\dagger} = \left(
\begin{array}{rr}
1 & -1\\
-1 & 1
\end{array} \right).
 \label{master-two}
\end{equation}
By introducing operators, $\hat{W}={\rm diag}(W_{1}, W_{2})$ and
$\hat{E}={\rm diag}(E_{1}, E_{2})$, so that
\begin{equation}
e^{\beta \hat{E}}=
\left( \begin{array}{cc}
\kappa_1 & 0 \\
0 & \kappa_2
\end{array} \right), \quad e^{-\beta \hat{W}}=
\left( \begin{array}{cc}
g_1 & 0 \\
0 & g_2
\end{array} \right),
\label{expeg}
\end{equation}
we can express the master operator, $\hat{H}$, in the form, given by Eq.~(\ref{master-operator-2}).

\subsection{Geometric properties of vector-potential ${\bm A}$ in toy model}
\label{toy-vector-potential}

The vector potential ${\bm A}$ can be derived in adiabatic limit by considering small deviation of the probability distribution from its instantaneous equilibrium,
${\bm \rho} = {\bm \rho}^{B} + \delta {\bm \rho}$, where ${\bm \rho}^{B}=Z^{-1}(e^{\beta E_1} \, ,e^{\beta E_2}) $, $Z=(e^{\beta E_1}+e^{\beta E_2})$. Substituting this into the master equation (\ref{partial}) and disregarding higher time-derivatives than the first one, we can find $\delta {\bm \rho}$ as a function of $d{\bm \rho}^{B}/dt$, i.e.
\begin{equation}
\delta \rho_1 = -\delta \rho_2 =\frac{-d \rho_1^{B}/dt}{(g_1+g_2)(\kappa_1+\kappa_2)}.
\label{deltrho-2state}
\end{equation}
We then find the average current using Eq.~(\ref{J-average-general}), which leads us to
\begin{equation}
\begin{array}{l}
J_1=g_1\kappa_1 \delta \rho_1 - g_1\kappa_2 \delta \rho_2,\\
\\
J_2=-g_2 \kappa_1 \delta \rho_1 + g_2 \kappa_2 \delta \rho_2.
\end{array}
\label{av-curr-2state}
\end{equation}
Due to Eqs.~(\ref{deltrho-2state}) and (\ref{av-curr-2state}), the differential of the total current, $d{\bar {\bm Q}}= {\bm J}dt$, can be written as
\begin{equation}    
d{\bar {\bm Q}} = {\bm A}^1d\rho_1^{B} + {\bm A}^2 d\rho_2^{B},
\label{geom-2state}
\end{equation}
where ${\bm A}^{1}=a(g_1+g_2)^{-1}(-g_{1},\,g_{2})$, ${\bm A}^{2}=(a-1){\bm A}^{1}$, and where
$a$ is arbitrary constant, which has no physical meaning. Its arbitrariness reflects the fact that, due to particle conservation, the vectors ${\bm A}^1$ and ${\bm A}^2$ are not independent. It is rather their difference,
\begin{equation}
{\bm A}^2-{\bm A}^1 =(g_{1}+g_{2})^{-1}\left(g_1,\,-g_2\right)
\label{ama-2state}
\end{equation}
that can be uniquely defined.

\begin{figure}
 \centerline{\includegraphics[width=2.8in]{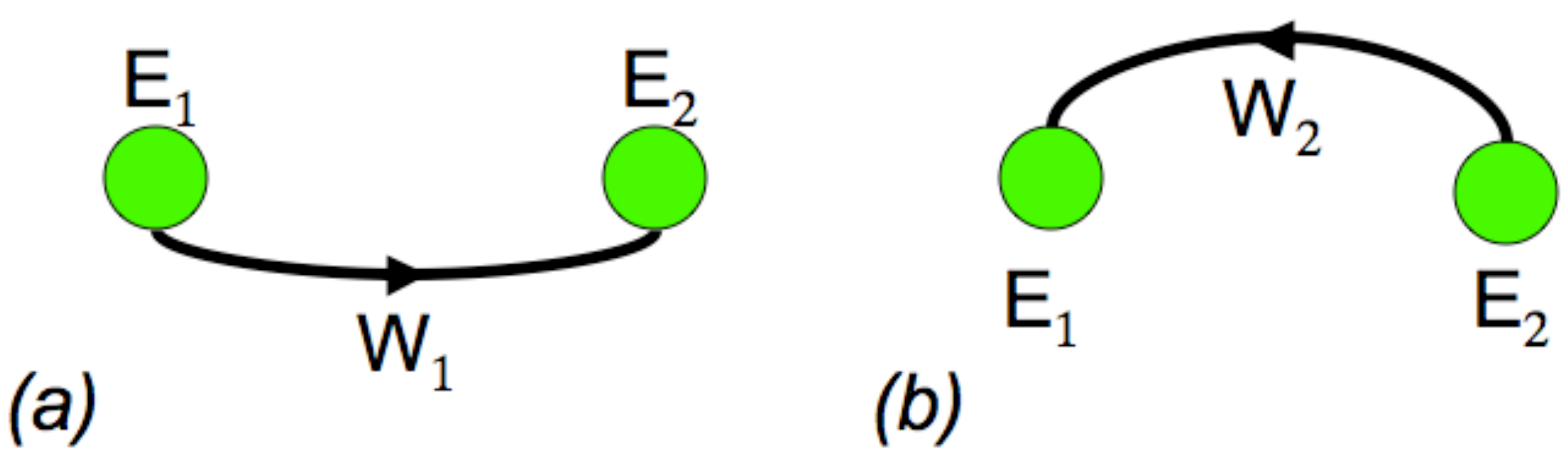}}
  \caption{ Minimal spanning trees for (a) $W_1<W_2$ and (b) $W_1>W_2$. At the low
    temperature limit, motion is restricted to be on a tree graph
    that, for toy model, consists only of one link and two nodes. The
    choice of the minimal spanning tree at each time interval depends on the relative
    strengths of barriers, e.g. interval (iv) corresponds to the tree
    (a) and interval (ii) corresponds to the tree (b).}
\label{2level-split}
 \end{figure}

To make parallels with the more general Eq.~(\ref{A-Kirchhoff-general}), note that
we can identify  ${\bm A}^2-{\bm A}^1$  by introducing the vector
\begin{equation}
{\bm \varrho}^B =(g_{1}+g_{2})^{-1}\left(g_1,\,g_2 \right)
\label{varrho}
\end{equation}
of Boltzmann weights with components on both of the graph's spanning trees, shown in Fig.~\ref{2level-split}. We can further introduce the vectors of currents,
\begin{equation}
{\bm Q}_{12}(\tilde{X}_1)=(1,0), \quad {\bm Q}_{12}(\tilde{X}_2) = (0,-1),
\label{veccur}
\end{equation}
which are generated by the paths that start at site 1 and end at site 2 staying, respectively, on the tree $\tilde{X}_1$ in Fig.~\ref{2level-split}(a) and on the tree $\tilde{X}_2$ in Fig.~\ref{2level-split}(b). Then,
\begin{equation}
{\bm A}^2-{\bm A}^1 = \sum_{\tilde{X}_i;\, i=1,2} {\bm Q}_{12}(\tilde{X}_i) {\bm \varrho}^{B}_{{\tilde{X}_i}},
\label{ama-2state-2}
\end{equation}
which shows that Eq.~(\ref{ama-2state}) is a special case of Eq.~(\ref{A-Kirchhoff-general}).

We would like also to discuss the analogy between the properties of the vector potential ${\bm A}$ and Kirchhoff's laws for electric circuits. The components of the vector ${\bm A}^2-{\bm A}^1$ can be viewed as  currents on the two links of the graph. An important observation is that those currents are the same as they would be if we consider the currents in a very simple electric problem, in which the links represent the resistances $R_{1}=g_{1}^{-1}$ and $R_{2}=g_{2}^{-1}$ for the first and the second link, respectively, and the nodes represent the junctions where a unit current enters (first node) and leaves (second node) the network, respectively. According to the Kirchhoff's laws, the drop of the voltages around the loop of the network must be zero, whereas the sum of the currents entering any node must be zero, which can be expressed as
\begin{equation}
\begin{array}{l}
J_{1}R_{1}  +J_{2}R_{2}= 0, \\
\\
1+J_{2}-J_{1}=0.
\end{array}
\label{k-2state}
\end{equation}
The solution of the above linear system [Eq.~(\ref{k-2state})] yields $(J_{1},\,J_{2}) =(g_1+g_2)^{-1}( g_1,\, -g_2)$, which coincides with Eq.~(\ref{ama-2state}). So we can identify the components of the vector ${\bm J}$ that solves the Kirchhoff's laws with components of the vector-potential,  ${\bm A}^{2}-{\bm A}^{1}$ , e.g., the latter also solves Eq.~(\ref{k-2state}). Note also that by using our notation, after we replace $(J_{1},\,J_{2})$ by the components of ${\bm A}^{2}-{\bm A}^{1}$, Eq.~(\ref{k-2state})
can be compactly  rewritten as
\begin{equation}
({\bm {\bm A}^2-{\bm A}^1},{\bm J}_{0})_g = 0, \quad \partial ({\bm A}^2-{\bm A}^1)_i = \delta^2_{i}-\delta^1_{i},
\label{notation-2state}
\end{equation}
where ${\bm J}_{0} = (1,1)$ is a unit conserved current that circulates around the loop in Fig.~\ref{2level-1}.

 \begin{figure}
\centerline{\includegraphics[width=2.2in]{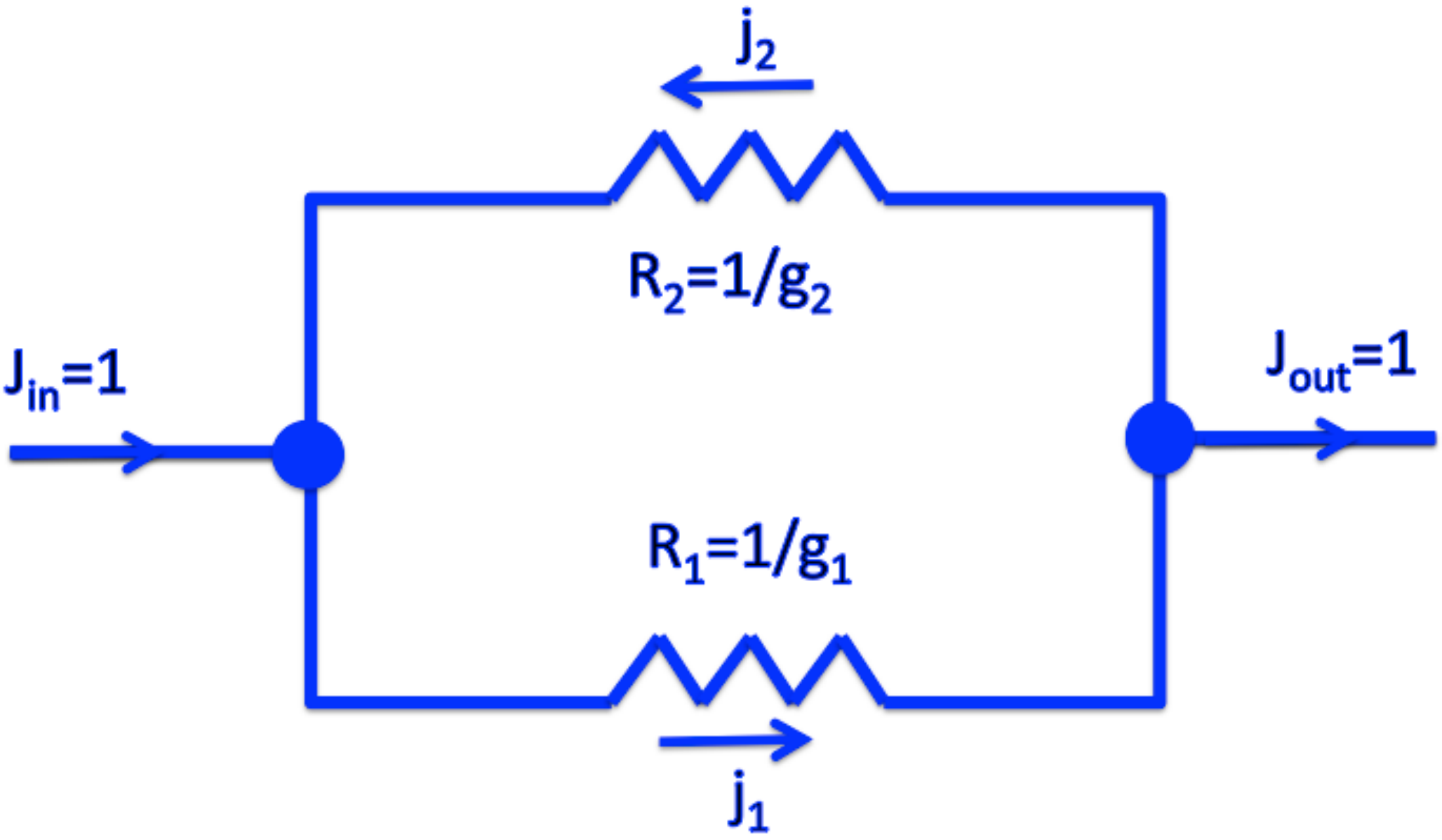}}
  \caption{ Resistance network that corresponds to the model in Fig.~\ref{2level-1}.
 }
 \label{resistances}
\end{figure}

\subsection{Current quantization in the low-temperature limit}
\label{toy-current-quantization}

For our toy model, the space $D_{X}$ of parameters with bad degeneracies is defined, simply, by the conditions $E_{1}=E_{2}$ and $W_{1}=W_{2}$. Any driving protocol that avoids this degeneracy space, hence, belongs to the space ${\cal M}_{X}$ of good parameters. For example, consider a protocol, given by $E_2=W_2=0$, $E_1(\tau)=\cos(2\pi \tau)$, $W_1(\tau)=\sin(2\pi \tau)$. One can distinguish four time
segments with distinct dynamical behavior patterns:

(i) for $\tau \in [- \varepsilon,\varepsilon] $, where $\varepsilon < \pi/2$, the path goes through the point $W_1=W_2$ but we always have $E_1>E_2$;

(ii) for $\tau \in [\varepsilon,\pi-\varepsilon]$, we have $W_1>W_2$;

(iii) for $\tau \in [\pi-\varepsilon,\pi+\varepsilon]$, we pass through the point $W_1=W_2$ but we always have $E_1<E_2$;

(iv) for $\tau \in [\pi+\varepsilon,2\pi-\varepsilon]$, we have $W_2>W_1$.

Consider stochastic dynamics on each of those segments in the low temperature limit. Whenever the barriers, $W_{1}$ and $W_{2}$ are different, when $\beta\to \infty$, we can omit in Eq.~(\ref{ama-2state}) the terms $e^{-\beta W_{1}}$ or $e^{-\beta W_{2}}$ if $W_1>W_2$, or $W_2>W_1$, respectively. This means that at the segments (ii) and (iv) ${\bm A}^2-{\bm A}^1$ is given by $(0,-1)$ and $(1,0)$, respectively.
At segments (i) and (iii) we encounter a situation when $ W_{1}=W_{2}$. However, at these
segments we have $E_{1}>E_{2}$ and $E_{2}>E_{1}$, which in the $\beta\to\infty$ limit means that the whole population is almost completely accumulated in the node $2$ and $1$, respectively. Therefore, the population vector practically does not change at these intervals, and, therefore, no significant average current is generated, according to Eq. (\ref{geom-2state}). Stated differently the current is generated at the segments (ii) and (iv) only. Note also that at the segments (ii) and (iv) the population difference $\rho_1-\rho_2$ changes from -1 to 1, and from 1 to -1, respectively.

Substituting the obtained values of ${\bm A}$ for the segments (ii) and (iv) into the integral expression for the average current [Eq.~(\ref{q-geom})] we find that current generated by the protocol (i)-(iv) is given by
\begin{equation}
{\bar {\bm Q}}({\bm s})=(1,1),
\label{pumpp-toy}
\end{equation}
i.e. the particle passes overall exactly one time through both links of the graph in the counterclockwise direction. Note also that this current is a conserved current in the sense that sum of its components entering any node equals the sum of components leaving this node.

The low-temperature average current, given by Eq.~(\ref{pumpp-toy}), can be interpreted using the following
{\it deterministic picture}. If the barrier  through the second link is higher, this link cannot be traversed and the motion is effectively restricted to the minimal spanning tree graph, as shown in Fig.~\ref{2level-split}(a). Similarly, when the first barrier is higher, the particle motion is restricted to the tree graph shown in Fig.~\ref{2level-split}(b).

At the start of the protocol (i)-(iv), the particle distribution is concentrated at site $2$ and all currents are exponentially suppressed. When energy $E_{1}$ becomes lower, the particle moves to site $1$ via
link $2$, and still stays in site $1$ while barrier $1$ becomes preferable for a transition and, therefore the particle returns to site $2$ via link $1$, resulting in quantized current given by Eq.~(\ref{pumpp-toy}). This picture, of course, is oversimplified because we apply the adiabatic limit {\it before} the low-temperature counterpart, so that the true motion is stochastic rather than deterministic. In fact, a particle can make, and actually makes a large number random transitions through both links during the cycle duration and quantization that we discuss appears only on average. Nevertheless, the deterministic picture
correctly predicts the average current.

\subsection{Winding index}
\label{toy-winding}

\begin{figure}
 \centerline{\includegraphics[width=2.4in]{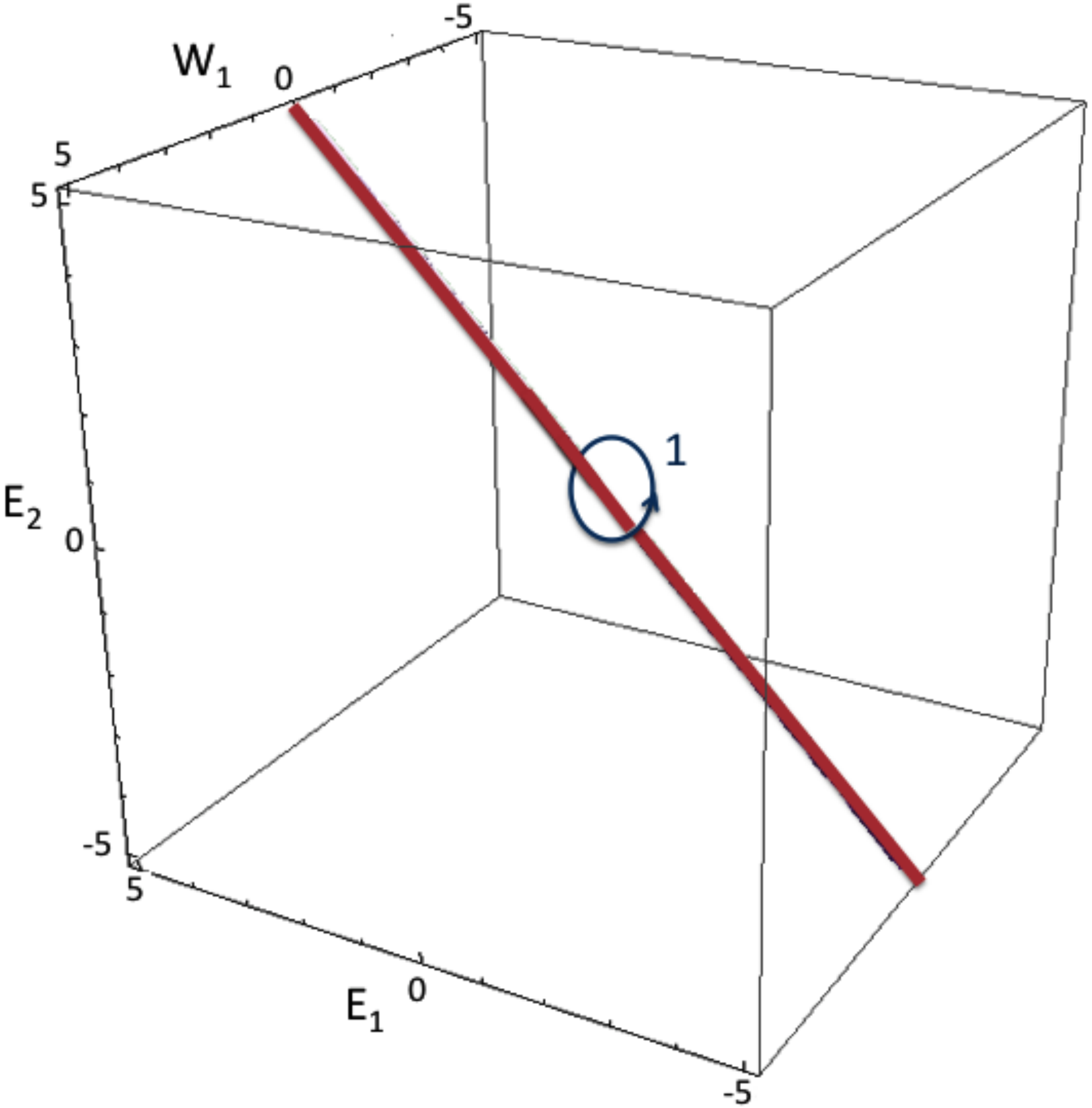}}
  \caption{ Space of the parameters with bad degeneracies (purpur) and a contour in the space  of good parameters (blue) that winds around the component of the bad parameter space. The winding number that corresponds to this contour is 1. Parameter $W_2$ is set to zero. }
 \label{pic-thread}
\end{figure}

The parameter space $\widetilde{{\cal M}}_{X}$ for our toy model is $4$-dimensional. The space $D_{X}$ of bad parameters is $2$-dimensional, being defined by two constraints, $E_{1}=E_{2}$ and $W_{2}=W_{2}$.
To visualize the structure of $D_{X}$, we fix  one of the barrier heights, $W_{2}=0$. The space $D_{X}$ of bad parameters is represented by a straight purple line in Fig.~\ref{pic-thread}. In the adiabatic low-temperature limit any driving protocol that encloses this line produces an integer-valued current, given by the protocol winding number around this line, i.e., the number of times the curve in the good parameter space that represents the protocol winds around the straight line that represents $D_{X}$.  There is an analogy between the above topological representation of pump currents and the Aharonov-Bohm effect in quantum mechanics. In the latter, the phase of the electronic wave function changes upon enclosing a quasi-one-dimensional solenoid with a magnetic field by an amount proportional to the total flux of the field inside the solenoid. This analogy follows from the similarity of the expression for pump current and the geometric (Berry) phase in quantum mechanics that has been discussed previously in the literature \cite{westerhoff-86,sinitsyn-07epl,astumian-07pnas,sinitsyn-09review,ohkubo-08jstat}.

To make the above analogy more explicit, as well as to understand better the nature of the low-temperature limit we consider a contour that encloses the bad parameter space only once, as shown in Fig.~\ref{pic-thread}. By applying the Stokes theorem to Eq.~(\ref{q-geom}), we can express  the corresponding pump current, ${\bm Q}({\bm s})$, generated at arbitrary finite temperature, in terms of an integral over a surface, ${\cal A}$, spanned on the contour, ${\bm s}$ that represents the driving protocol
\begin{eqnarray}
\label{toy-Stokes} {\bm Q}({\bm s})&=&\int_{{\cal A}}d{\bm
A}\wedge d{\bm\rho} \\ &=&\sum_{j\alpha}\int_{{\cal A}}{\bm F}^{\alpha j}({\bm E},{\bm W})dE_{j}\wedge dW_{\alpha},
\end{eqnarray}
where, in our toy model, $i,\alpha=1,2$, and we define the curvature ${\bm F}$ of the vector potential ${\bm A}$ in a standard way
\begin{eqnarray}
\label{toy-F-adiabatic} {\bm F}^{\alpha j}({\bm E},{\bm W})=\sum_{k=1,2}\frac{\partial{\bm A}^{k}({\bm W})} {\partial W_{\alpha}}\frac{\partial\rho_{k}({\bm E})}{\partial E_j}.
\end{eqnarray}
Using the mentioned above analogy with the phase of a quantum wavepacket we can view the curvature ${\bm F}$ as a ``magnetic field'' in the parameter space \cite{sinitsyn-09jcp}. For low enough temperatures this magnetic field is concentrated along the bad parameter space, as shown in  Fig.~\ref{pic-thread}. The total flux of this field equals $1$. Decreasing temperature in the model does not change the total flux, while forcing the field to be concentrated in a narrow tubular neighborhood of the straight line that represents the bad parameter space. In the $\beta\to \infty$ limit this field is restricted to an infinitely narrow $1$D solenoid along the bad parameter space. This leads to the above mentioned analogy with the Aharonov-Bohm effect.

\section{Integer quantization}
\label{integer-Q}

In this section,  we return to  the arbitrary graph geometry and study the low-temperature adiabatic limit $\bar{{\bm Q}}({\bm s})=\lim_{\beta\to\infty}\lim_{\tau_{D}\to\infty}{\bm Q}({\bm s};\beta,\tau_{D})$ in the situation when the driving protocol stays within the subspace ${\cal M}_{X}$ of good parameters for integer quantization (see subsection~\ref{good-bad-parmeters} for the definition of ${\cal M}_{X}$). 
Let us introduce notation, $k\otimes\tilde{X}$, which means the unit vector in combined population-current space. This vector has unit components on the node $k$ and links of the tree  $\tilde{X}$ and has zero other components. 
 The sum-over-trees expression for the adiabatic limit ${\bm Q}({\bm s};\beta)$ of the average current [Eqs. (\ref{q-explicit-adiabatic}) and (\ref{gamma-distr-adiabatic})] is very convenient for taking the low-temperature limit, which boils down to identifying the low-temperature limit of ${\bm\gamma}=\sum_{k,\tilde{X}}\gamma_{k\tilde{X}}k\otimes\tilde{X}$
\begin{eqnarray}
\label{q-adiabatic-low-T} \bar{{\bm Q}}({\bm s})=\sum_{k,\tilde{X}}{\bm Q}^{k\tilde{X}}\bar{\gamma}_{k\tilde{X}}({\bm s}), \; \bar{\gamma}_{k\tilde{X}}({\bm s})=\lim_{\beta\to\infty}\gamma_{k\tilde{X}}({\bm s};\beta),
\end{eqnarray}
provided the $\beta\to\infty$ limit in Eq.~(\ref{q-adiabatic-low-T}) does exist. In this section we explicitly calculate the low-temperature limit $\bar{{\bm\gamma}}(\bm s)$ in the case when the driving protocol fully lies in the space ${\cal M}_{X}$ of good parameters for integer quantization, and show, in particular, that the components $\bar{\gamma}_{k\tilde{X}}$ are all integer, which implies integer quantization of average current in the setting under consideration.

\subsection{Maximal-weight and minimal spanning trees}
\label{minimal-spanning}

In the adiabatic limit, the distribution ${\bm\gamma}({\bm s})$ is explicitly expressed in terms of the Boltzmann distributions ${\bm\rho}^{B}$ and ${\bm\varrho}^{B}$ in the node and spanning tree spaces, respectively. In the case when there is no degeneracy of the lowest node energy $E_{k}$ the Boltzmann distribution has a well-defined low-temperature limit $\bar{\rho}_{j}^{B}=\delta_{jk}$, i.e. the population is concentrated in the lowest-energy node $k$. The low-temperature limit of the tree Boltzmann distribution ${\bm\varrho}^{B}$ under the condition of no degeneracy between any two barrier heights is less obvious. The limit can still be identified by  introducing a notion of the {\it minimal spanning tree} $\tilde{X}_{{\bm W}}$, and its two properties: (i) the tree $\tilde{X}_{{\bm W}}$ is non-degenerate, and minimizes the energy $W(\tilde{X})$, (ii) the tree $\tilde{X}_{{\bm W}}$ is totally determined by the ordering of the barrier heights, the actual values of the barrier heights within a given ordering is irrelevant. Property (i) completely identifies the low temperature limit $\bar{\varrho}_{\tilde{X}}^{B}= \delta_{\tilde{X}\tilde{X}_{{\bm W}}}$ of the tree Boltzmann distribution in the case of non-degenerate barrier data. Property (ii) justifies the  notation $\tilde{X}_{[{\bm W}]}$, where $[{\bm W}]$ denotes the equivalence class of the barrier data with the same ordering of the barrier heights.

The minimal spanning tree $\tilde{X}_{{\bm W}}$ of $X$, associated with non-degenerate barrier data ${\bm W}$ is a subgraph, constructed by sequentially removing the links with the highest possible barriers such that the remaining graph remains connected. Eventually, this process ends up with a tree (i.e. a graph without loops), referred to as the minimal spanning tree.


The minimal spanning tree has an obvious, yet very important property that can be actually used as its alternative definition, and, therefore, referred to as the {\it characterizing property} of a minimal spanning tree: For each withdrawn link (i.e., a link that does not belong to $\tilde{X}_{{\bm W}}$) $\alpha$ with $\partial\alpha=\{i,j\}$, the barrier $W_{\alpha}$ is higher than any of the barriers associated with the links of the path $l_{ij}(\tilde{X}_{{\bm W}})$. In particular, the physical meaning of the introduced current vector, ${\bm Q}(\tilde{X}_{{\bm W}})$ that corresponds to a path path $l_{ij}(\tilde{X}_{{\bm W}})$, is that it is generated when a particle moves from node $i$ to node $j$ on the graph $X$, avoiding the highest barriers.

Property (ii) immediately follows from the above explicit construction of $\tilde{X}_{\bm W}$ that uses the ordering of the barrier heights only. Property (i) is demonstrated in appendix~\ref{min-span-details}. In this subsection we just present a physical argument in support of property (i). According to the sum-over-trees representation the population redistribution in the $\beta\to\infty$ is restricted to the spanning tree of the maximal weight, i.e., of the minimal value of $W(\tilde{X})$. On the other hand, random motion of the particle tends to avoid the links with the highest barriers, and, therefore, should be restricted to the minimal spanning tree $\tilde{X}_{{\bm W}}$, which supports the statement of property (i).

\subsection{Good parameters, minimal spanning trees, and quantized currents}
\label{good-param-min-span}


Current generation in the low-temperature adiabatic limit can be interpreted using a simple and clear intuitive picture. When the protocol goes over the subspace $U_{0}\subset {\cal M}_{X}$, when the lowest
node energy is non-degenerate, the particle sits in the unique lowest-energy node, and no current is generated at all. Current can be generated only during time intervals when at least two lowest node energies are close to their degeneracy. However, in this case, by the definition of good parameters, the barrier data
${\bm W}$ is non-degenerate (i.e., the protocol goes through $U_{1}\subset {\cal M}_{X}$), and population redistribution between any two nodes $i$ and $j$, to avoid the highest barriers, is restricted to the tree $\tilde{X}_{W}$, and, therefore, produces the current proportional to ${\bm Q}_{ij}(\tilde{X}_{{\bm W}})$.



To make use of the above argument, we partition the time period of a driving protocol into a set of small enough segments, so that for each segment we either encounter only the minimal well depth degeneracy or just the potential barrier degeneracy (this can always be achieved due to the well-known argument of van Kampen). We refer to them as to $0$- and $1$-segments, respectively. If necessary, we merge the consecutive segments of the same type to make the $0$- and $1$-segments alternating. In the $\beta \rightarrow \infty$ limit, no current is generated at any $1$-segment, since the populations is concentrated at the node with the lowest energy value, while nonzero average pump current can occur only when the state probability vector changes \cite{jarzynski-08prl}. Since the $1$-segment ends also belong to $0$-segments, in the $\beta\to\infty$ limit the populations at the beginning and at the end of a 1-segment, labeled $a$, are concentrated in well-defined nodes, denoted by $k_{a}$ and $k'_{a}$, respectively, and determined by the neighboring $0$-segments.

As already argued above, when the particle moves from one node to another to generate current, it
avoids the highest barriers i.e. it moves along $\tilde{X}_{{\bm W}}$, which is associated with
non-degenerate barrier data ${\bm W}$ . Since the paths with non-vanishing probabilities from  $k_{a}$ to $k'_{a}$ belong to a tree $\tilde{X}_{{\bm W}_{a}}$, the total current ${Q}^{a}$, passing during the time of the 1-segment $a$ has values 1 on any link on the shortest path $l_{k_{a}k'_{a}}(\tilde{X}_{{\bm W}_{a}})$
and zero values on the other links. Finally, the current per cycle is generated by the concatenation of the
consecutive paths $l_{k_{a}k'_{a}}(\tilde{X}_{{\bm W}_{a}})$ that correspond to the $a$-segments. This explicitly identifies the generated integer-valued average current per cycle,
\begin{eqnarray}
\label{Q-double-limit} \bar{{\bm Q}}({\bm s})=\sum_{a}{\bm
Q}_{k_{a}k'_{a}}(\tilde{X}_{{\bm W}_{a}}),
\end{eqnarray}
where the summation runs over all $1$-segments, and, combined with the definition of ${\bm Q}_{ij}(\tilde{X}_{{\bm W}})$, Eq.~(\ref{Q-double-limit}) yields integer-valued $\bar{{\bm Q}}({\bm s})$. Since, by construction, the $1$-segments are alternating with the $0$-segments, we have
for consecutive $1$-segments $k'_{a}=k_{a+1}$, which implies $\partial\sum_{a}{\bm
Q}_{k_{a}k'_{a}}(\tilde{X}_{{\bm W}_{a}})=0$, or equivalently $\bar{{\bm Q}}({\bm s})$ is a conserving integer-valued (quantized) current.

We have derived the expression for $\bar{{\bm Q}}({\bm s})$ [Eq.~\ref{Q-double-limit})] using plausible and transparent physical argument. We are now in a position to sketch a rigorous proof of the integer PQT, or more specifically of the validity of Eq.~(\ref{Q-double-limit}). The sum-over-trees formula [Eq.~(\ref{q-adiabatic-low-T})] allows us to identify the low-temperature limit, $\bar{{\bm\gamma}}({\bm s})$, of the joint distribution ${\bm\gamma}$. We start with a formal notation for the aforementioned partitioning of the driving protocol time period $[0,1]/\{0,1\}=S^{1}=\bigcup_{a}(I_{a}\cup\tilde{I}_{a})$, where $\tilde{I}_{a}=[\tau'_{a-1},\tau_{a}]$ and $I_{a}=[\tau_{a},\tau'_{a}]$ are the alternating $0$- and $1$-segments, respectively. We further recast the expression for the adiabatic, still finite-temperature limit of the joint distribution [Eq.~(\ref{gamma-distr-adiabatic})] as a sum of the segment contributions
\begin{eqnarray}
\label{gamma-adiabatic-segments} {\bm\gamma}({\bm s};\beta)&=&\sum_{a}\int_{\tau'_{a-1}}^{\tau_{a}}d\tau{\bm\varrho}^{B}\otimes\dot{{\bm\rho}}^{B} \nonumber \\ &+& \sum_{a}\int_{\tau_{a}}^{\tau'_{a}}d\tau{\bm\varrho}^{B}\otimes\dot{{\bm\rho}}^{B},
\end{eqnarray}
and apply integration by parts to the second term (sum over $1$-segments), resulting in
\begin{eqnarray}
\label{gamma-adiabatic-segments-2} {\bm\gamma}({\bm s};\beta)&=& \sum_{a}\bm{\varrho}^{B}({\bm s}(\tau'_{a});\beta)\otimes\bm{\rho}^{B}({\bm s}(\tau'_{a});\beta) \nonumber \\ &-& \sum_{a}\bm{\varrho}^{B}({\bm s}(\tau_{a});\beta)\otimes\bm{\rho}^{B}({\bm s}(\tau_{a});\beta) \nonumber \\ &+& \sum_{a}\int_{\tau'_{a-1}}^{\tau_{a}}d\tau{\bm\varrho}^{B}({\bm s}(\tau);\beta)\otimes\dot{{\bm\rho}}^{B}({\bm s}(\tau);\beta) \nonumber \\ &-& \sum_{a}\int_{\tau_{a}}^{\tau'_{a}}d\tau\dot{{\bm\varrho}}^{B}({\bm s}(\tau);\beta)\otimes{\bm\rho}^{B}({\bm s}(\tau);\beta).
\end{eqnarray}
Before applying the low-temperature limit to Eq.~(\ref{gamma-adiabatic-segments-2}) we would like to note that the expression is exact, and is valid for any partition of the time period into a set of alternating segments. In particular, it will be used in subsection~\ref{A-low-temp-rational} to identify the low-temperature limit for the rational quantization setting (permanent degeneracies).

The $\beta\to\infty$ limit of Eq.~(\ref{gamma-adiabatic-segments-2}) for the integer quantization setting under consideration is obtained using the following facts: (i) The parameter sets that correspond to the segment ends ${\bm s}(\tau_{a}),{\bm s}(\tau'_{a})\in U_{0}\cap U_{1}$ represent non-degenerate lowest-energy and barrier data, and, therefore, the node and tree Boltzmann distributions in the contact (the first two) terms in Eq.~(\ref{gamma-adiabatic-segments-2}) have well-defined limits with the populations concentrated in the lowest-energy node and minimal spanning tree, respectively; (ii) the minimal spanning tree $\tilde{X}_{{\bm W}(\tau)}$ is the same for all times that belong to a $1$-segment $I_{a}$, so that a notation $\tilde{X}_{a}$ makes perfect sense; (iii) the integral terms (the last two) in Eq.~(\ref{gamma-adiabatic-segments-2}) vanish in the $\beta\to\infty$ limit. Based on these properties the low-temperature limit of the joint distribution is identified as
\begin{eqnarray}
\label{gamma-adiabatic-segments-LT} \bar{{\bm\gamma}}({\bm s})= \sum_{a}\tilde{X}_{a}\otimes(k'_{a}-k_{a}),
\end{eqnarray}
and Eq.~(\ref{Q-double-limit}) is obtained by applying the bilinear form ${\bm Q}$ with the components ${\bm Q}^{k\tilde{X}}$ to Eq.~(\ref{gamma-adiabatic-segments-LT}), making use of the property ${\bm Q}(\tilde{X}\otimes(k'-k))={\bm Q}_{kk'}(\tilde{X})$.

Property (i) is obvious. Property (ii) follows from the connected nature of a segment topology and the fact that $\tilde{X}_{\bm W}=\tilde{X}_{[{\bm W}]}$. Property (iii) is established by representing
\begin{eqnarray}
\label{rhoB-dot-gradients} \dot{{\bm\rho}}^{B}=\dot{{\bm E}}\cdot\partial_{{\bm E}}{\bm\rho}^{B}, \;\;\; \dot{{\bm\varrho}}^{B}=\dot{{\bm W}}\cdot\partial_{{\bm W}}{\bm\varrho}^{B},
\end{eqnarray}
[where ${\bm W}$ is considered as a vector in the space over the spanning trees with the components $W_{\tilde{X}}=W(\tilde{X})$], and noting that $\dot{{\bm E}}$, $\dot{{\bm W}}$, ${\bm\rho}^{B}$, and ${\bm\varrho}^{B}$ are bound, and the gradient of a Boltzmann distribution tends to zero in the $\beta\to\infty$ limit in the case of non-degenerate lowest-energy state (node or tree). The last property can be verified by obvious and straightforward estimates.

We conclude this subsection with emphasizing that the presented derivation not only demonstrates integer quantization of pumped currents in the low-temperature adiabatic limit, but rather provides an algorithmic way of computing the integer-valued current, given the driving protocol. The integer quantization formula
(\ref{Q-double-limit}) is, obviously, a special case of the AF (\ref{Q-averaging55}).

\subsection{Deterministic picture for integer quantization}
\label{integer-deterministic}

In this subsection we illustrate the construction of integer quantized current $\bar{{\bm Q}}({\bm s})$ as a sum of piecewise integer currents $ {\bm Q}_{k_{a}k'_{a}}(\tilde{X}_{{\bm W}_{a}})$, defined on minimal
spanning trees (see Fig.~\ref{network-steps}). The path in the control parameter space can be split into overlapping intervals. Intervals of the first type have an unchanging node with the lowest energy, while intervals of the second type correspond to transitions between nodes but having the minimal spanning tree unchanged.

For a graph in Fig.~\ref{network-steps}(a), we illustrate the deterministic evolution for some periodic protocol. Orange arrows indicate the path that dominates the average currents at low temperature. A particle starts at the minimal node 1 having the largest barriers along the links connecting nodes $\{3,4\}$, $\{1,6\}$ and $\{2,4\}$. By the time when this configuration of largest barriers changes the node with the lowest energy becomes the node 3. The average current during this time interval can be computed by assuming that a particle goes deterministically from the node 1 to the node 3 via the path indicated by  orange arrows in Fig.~\ref{network-steps}(b). After this, the protocol proceeds with a time interval during which a particle stays at the node 3 until the barrier configuration is such that the minimal spanning tree is the one shown in Fig.~\ref{network-steps}(c). We then repeat our arguments until the protocol completes one period and the particle returns to the node 1.  After the protocol illustrated in Fig.~\ref{network-steps}(b)-(d), a unit current will be generated through links $\{1,2\}$], $\{2,3\}$, $\{3,4\}$, $\{4,5\}$, $\{5,6\}$, $\{6,1\}$, and zero current will pass on average through links $\{1,4\}$ and $\{2,4\}$.

\section{Average currents and rational quantization}
\label{average-Q-rational}

In this section, we study the low-temperature adiabatic limit of $\bar{{\bm Q}}({\bm s})$ in the situation with permanent degeneracy (i.e., rational quantization setting). This is achieved by applying the low temperature $\beta\to\infty$ limit to the adiabatic limit ${\bm Q}({\bm s};\beta)$ of the average current [Eqs.~(\ref{q-explicit-adiabatic}) and (\ref{gamma-distr-adiabatic})], and ultimately to Eq.~(\ref{gamma-adiabatic-segments-2}). We will further derive the averaging formula [Eq.~(\ref{Q-averaging55})] that expresses $\bar{{\bm Q}}({\bm s})$ as a result of equal-weight averaging of a properly chosen set of integer currents.

\subsection{Minimal spanning trees in the degenerate case}
\label{graphs-W}

The minimal spanning tree, $\tilde{X}_{{\bm W}}$, associated only with non-degenerate barrier data ${\bm W}$, as established in subsection~\ref{minimal-spanning} and appendix~\ref{min-span-details} maximizes its Boltzmann weight, and in the low-temperature $\beta\to\infty$ limit dominates it.

In the case of degeneracy of two or several barriers, we can instead identify a set of spanning trees that play a similar role to the role that minimal spanning trees play for graphs without barrier degeneracies. We will still refer to them as minimal spanning trees, having in mind that in the degenerate situation under discussion a minimal spanning tree is generically not unique.

A minimal spanning tree can be obtained from the original graph by the following procedure. We remove links
with the highest barriers one-by-one, and stop when the removal of any additional link of the chosen group breaks the graph into disjoint components. We then repeat the procedure with set of next-highest links and so on until we end up with a tree. In a generic situation a minimal spanning tree is not unique, since its construction explicitly depends on a particular choice of the withdrawn links within any group of degenerate links.

For a non-degenerate (in the sense of rational quantization setting, i.e., no degeneracies in addition to permanent degeneracies) barrier data ${\bm W}$ the set of minimal spanning trees is denoted by $\tilde{{\cal X}}_{{\bm W}}$. By the construction of the minimal spanning trees, the set $\tilde{{\cal X}}_{{\bm W}}$ depends on the ordering of the barrier groups heights, rather than their actual values. Therefore, similar to the integer quantization setting, a notation $\tilde{{\cal X}}_{[{\bm W}]}$ is in place. We will use a notation $n_{{\bm W}}(X)={\rm card}(\tilde{{\cal X}}_{{\bm W}})$ for the number of minimal spanning trees of the graph $X$ with given barrier data ${\bm W}$.

\subsection{Properties of minimal spanning trees and associated graphical structures}
\label{min-span-structures}

The crucial role of the sets $\tilde{{\cal X}}_{{\bm W}}$ of minimal spanning trees for the low-temperature limit is determined by their following properties: (i) The energies $W(\tilde{X})$ of all $\tilde{X}\in \tilde{{\cal X}}_{{\bm W}}$ are identical, and (ii) the energy $W(\tilde{Y})$ of any spanning tree $\tilde{Y}$ that does not belong to $\tilde{{\cal X}}_{{\bm W}}$ is strictly larger then the energy of the spanning trees, and, therefore, the minimal spanning trees dominate the Boltzmann distribution ${\bm\varrho}^{B}$ in the low-temperature limit.

All minimal spanning trees can be described using the following simple construction. For given barrier data ${\bm W}$, let us order the barrier values $W_{\mathfrak{A}_{1}}>\ldots>W_{\mathfrak{A}_{m}}$ and withdraw the links $\alpha\in\mathfrak{A}_{1}$ with the highest barriers $W_{\alpha}=W_{\mathfrak{A}_{1}}$. The resulting graph is generally disconnected and is represented by a disjoint union of its connected components, referred to as children of $X$, with $X$ naturally treated as their parent. We apply the above procedure to the children repeatedly, terminating when a child graph is a tree. All subgraphs obtained on the described way, including $X$ are referred to as the {\it derived subgraphs} of $X$. Let ${\cal G}_{X}({\bm W})$ be a tree graph whose nodes are the derived subgraphs of $X$, and the links represent the parent-child relation. Obviously the root of ${\cal G}_{X}({\bm W})$ is given by $X$, whereas all leaves are represented by trees.

With any non-tree derived subgraph we can associate its reduced counterpart, referred to as a {\it reduced derived subgraph} by contracting to a point all links except for the ones with the maximal barrier value. Naturally the nodes of a reduced subgraph are represented by the children of the original subgraph, with the links given by the links of the original subgraph with the maximal barrier value. The reduced subgraph of a tree [i.e. a leaf of ${\cal G}_{X}({\bm W})$] is defined as a result of contracting the tree to a point. Let $\tilde{{\cal G}}_{X}({\bm W})$ be the tree obtained by replacing the nodes of ${\cal G}_{X}({\bm W})$, represented by derived subgraphs with their reduced counterparts. Note that all links of any reduced derived subgraph have the same barrier values, and the set of links $X_{1}$ of our original network $X$ is given by a disjoint union of the link sets of the reduced derived subgraphs and the links of the leaves of ${\cal G}_{X}({\bm W})$ (i.e. tree-like derived subgraphs).

Naturally, any minimal spanning tree $\tilde{X}\in \tilde{{\cal X}}_{{\bm W}}$ is totally determined by a particular choice of a spanning tree for any reduced derived subgraph of $X$. The links of a minimal spanning tree are represented by a disjoint union of the link sets of the chosen spanning trees in the reduced derived graphs, and the link sets of the tree-like derived subgraphs [leaves of ${\cal G}_{X}({\bm W})$].

Property (i) of minimal spanning trees follows from the general fact that all spanning trees of a graph have the same number of links, applied to the chosen spanning trees of the reduced derived subgraphs of $X$. Therefore, any minimal spanning tree has the same number of links of any given barrier value, which implies the same value of $W(\tilde{X})$ for all $\tilde{X}\in \tilde{{\cal X}}_{{\bm W}}$. Property (ii) is validated in appendix~\ref{min-span-details}.

The explicit description of the minimal spanning trees presented in this subsection also provides  a convenient expression for the number $n_{{\bm W}}(X)$ of the minimal spanning trees as the product of the spanning trees numbers over all reduced derived subgraphs of $X$ [i.e., nodes of $\tilde{{\cal G}}_{X}({\bm W})$].

We conclude this subsection with noting that the presented construction based on reduced derived subgraphs has a transparent physical meaning. Since in the $\beta\to\infty$ limit population relaxation within child subgraphs occurs much faster than the population exchange between them, the reduced derived subgraphs represent the hierarchy of the relaxation processes time scales.

\subsection{Low-temperature limit for the current and minimal spanning trees}
\label{A-low-temp-rational}

Now we show that in the case with permanent degeneracy, for any good protocol ${\bm s}$, i.e. when ${\bm s}(\tau)$ stays within ${\cal M}_{X}$ at all times, there is a well-defined low-temperature adiabatic limit $\bar{{\bm Q}}({\bm s})= \lim_{\beta\to\infty}\lim_{\tau_{D}\to\infty}{\bm Q}({\bm s};\beta,\tau_{D})$, so that $\bar{{\bm Q}}({\bm s})$ is rational-valued, and present explicit ways to evaluate this limit. Our derivation is very similar to the one for the integer case, presented in subsection~\ref{good-param-min-span}, and is based on identification of the low temperature limit of Eq.~(\ref{gamma-adiabatic-segments-2}), where the differences reflect just the differences in the low-temperature limiting behavior of the node and tree Boltzmann distribution ${\bm\rho}^{B}({\bm x};\beta)$ and ${\bm\varrho}^{B}({\bm x};\beta)$ for ${\bm x}\in U_{0}$ (no energy data degeneracy) and ${\bm x}\in U_{1}$ (no barrier data degeneracy), respectively. The latter have the form
\begin{eqnarray}
\label{rho-B-low-temp} \lim_{\beta\to\infty}{\bm\rho}^{B}({\bm E};\beta)=\nu_{{\bm E}}^{-1}\sum_{k\in {\cal K}_{\bm E}}k,
\end{eqnarray}
for the node Boltzmann distribution, where ${\cal K}_{{\bm E}}\in {\cal X}_{0}$ denotes the subset of nodes $k$ with the lowest value of $E_{k}$, and $\nu_{{\bm E}}\equiv {\rm card}({\cal K}_{{\bm E}})$ is the corresponding number of nodes, and
\begin{eqnarray}
\label{A-Kirchhoff-low-temp} \lim_{\beta\to\infty}{\bm\varrho}^{B}({\bm W};\beta)=n_{{\bm W}}^{-1}\sum_{\tilde{X}\in \tilde{{\cal X}}_{\bm W}}\tilde{X}.
\end{eqnarray}
for the tree Boltzmann distribution.

Following our derivation of the integer case we partition the time period of the driving protocol into a set of alternating $0$- and $1$-segments exactly as described in subsection~\ref{good-param-min-span} and apply the low-temperature limit to Eq.~(\ref{gamma-adiabatic-segments-2}). By exactly the same reasons as in the integer case, the integral terms vanish in the $\beta\to\infty$ limit. We further note that, since the ends ${\bm s}(\tau_{a}),{\bm s}(\tau_{a}')\in U_{0}\cap U_{1}$ for all $a$, the low-temperature limits [Eqs.~(\ref{rho-B-low-temp}) and (\ref{A-Kirchhoff-low-temp})] apply to the contact terms in Eq.~(\ref{gamma-adiabatic-segments-2}). Therefore, denoting for brevity ${\cal K}_{a}= {\cal K}_{{\bm E}(\tau_{a})}$, ${\cal K}'_{a}= {\cal K}_{{\bm E}(\tau'_{a})}$, $\nu_{a}= \nu_{{\bm E}(\tau_{a})}$,  and $\nu'_{a}= \nu_{{\bm E}(\tau'_{a})}$ we arrive at
\begin{eqnarray}
\label{J-1-segment-limit-rational-2} \bar{{\bm\gamma}}({\bm s})&=&\sum_{a}\frac{1}{n_{{\bm W}_{a}}\nu_{a}\nu_{a}'} \nonumber \\ &\times& \sum_{k\in{\cal K}_{a}}\sum_{k'\in{\cal K}'_{a}}\sum_{\tilde{X}\in \tilde{{\cal X}}_{{\bm W}_{a}}}\tilde{X}\otimes (k'-k).
\end{eqnarray}

Finally, applying the bilinear form ${\bm Q}$ with the components ${\bm Q}^{k\tilde{X}}$ to Eq.~(\ref{J-1-segment-limit-rational-2}), exactly in the same way as for the integer case, we obtain
\begin{eqnarray}
\label{Q-double-limit-rational} \bar{{\bm Q}}({\bm s})=\sum_{a}\frac{1}{n_{{\bm W}_{a}}\nu_{a}\nu_{a}'}\sum_{k\in{\cal K}_{a}}\sum_{k'\in{\cal K}'_{a}}\sum_{\tilde{X}\in \tilde{{\cal X}}_{{\bm W}_{a}}}{\bm Q}_{kk'}(\tilde{X}),
\end{eqnarray}
and, combined with the definition of ${\bm Q}_{ij}(\tilde{X})$, Eq.~(\ref{Q-double-limit-rational}) yields rational-valued $\bar{{\bm Q}}({\bm s})$.

\subsection{Averaging formula for rational quantized currents}
\label{averaging-formula}

The expression [Eq.~(\ref{Q-double-limit-rational})] for the low-temperature adiabatic limit of the pumped current,  not only demonstrates the rational nature of the pumped currents, but also provides an algorithm for its computation. Yet, the aforementioned expression represents the pumped current as a rational superposition of integer currents generated at the $1$-segments, so that the decomposition can depend on a specific choice of partitioning the time period into the set of alternating $0$- and $1$-segments. In this subsection we derive an expression for $\bar{{\bm Q}}({\bm s})$ in the form of a rational superposition of integer-valued currents, defined globally, i.e., for the complete period of the driving protocol.

Each integer-valued component in the superposition is associated with a global degeneracy {\it resolution}. To introduce a degeneracy resolution we denote by $[{\bm E}]$ and $[{\bm W}]$ for any non-degenerate energy ${\bm E}$ and barrier ${\bm W}$ data the connected components of ${\bm E}$ and ${\bm W}$ in $U_{0}\subset\tilde{{\cal M}}_{X}$ and $U_{1}\subset\tilde{{\cal M}}_{X}$, respectively. Stated differently, $[{\bm E}]$ is fully determined by the lowest-energy subset ${\cal K}_{{\bm E}}$, and we can consider $[{\bm E}]\in {\cal X}_{0}$; naturally $[{\bm W}]$ is fully determined by a linear ordering of ${\cal X}_{1}$, so that we can consider $[{\bm W}]\in{\cal O}({\cal X}_{1})$, with ${\cal O}({\cal X}_{1})$ being the set of permutations (or equivalently linear orderings) of ${\cal X}_{1}$. A global degeneracy resolution $({\bm k},\tilde{{\bm X}})$ associates a node $k_{{\bm E}}\in {\cal K}_{{\bm E}}$ with any $[{\bm E}]\in {\cal X}_{0}$ and an optimal spanning tree $\tilde{X}_{{\bm W}}\in \tilde{{\cal X}}_{{\bm W}}$ with any $[{\bm W}]\in {\cal O}({\cal X}_{1})$. Obviously, the number $n_{X}$ of possible degeneracy resolutions is given by
\begin{eqnarray}
\label{n-X} n_{X}=\prod_{[{\bm E}]\in {\cal X}_{0}}\nu_{{\bm E}}\prod_{[{\bm W}]\in{\cal O}({\cal X}_{1})}n_{{\bm W}}.
\end{eqnarray}

To derive the AF, we consider Eq~(\ref{Q-double-limit-rational}) and we add the identity resolution,
\begin{equation}
\frac{ \nu_{a} \nu_{a}' n_{{\bm W}_{a}     }} {\nu_E^2 n_{X}} \sum_{b \ne a}\sum_{k_{b}\in{\cal K}_{b}}\sum_{k'_{b}\in{\cal K}'_{b}}\sum_{\tilde{X}\in \tilde{{\cal X}}_{{\bm W}_{b}}}1 = 1,
\label{id-res}
\end{equation}
to each term in the sum over time intervals $a$. Using that $\sum_{[{\bm E}]\in {\cal X}_{0}}1= \nu_{E}$, and that the sum over all possible identity resolutions, $\sum_{({\bm k},\tilde{{\bm X}})}$, is given by $\sum_{({\bm k},\tilde{{\bm X}})}  \equiv \sum_{[{\bm E}]\in {\cal X}_{0}}  \sum_{[{\bm W}]\in{\cal O}({\cal X}_{1})} $ we obtain
\begin{eqnarray}
\label{Q-double-limit-rational-1} \bar{{\bm Q}}({\bm s})=\sum_{a}\frac{1}{n_{X}} \sum_{({\bm k},\tilde{{\bm X}})}{\bm Q}_{k_{a}k'_{a}}(\tilde{X}).
\end{eqnarray}
Since $\frac{1}{ n_{X}}$ is a constant, we can interchange positions of the sums in (\ref{Q-double-limit-rational-1}). We  then collect the piecewise currents  ${\bm Q}_{k_{a}k'_{a}}(\tilde{X})$ in groups. The sum of the currents in each group represents some integer valued {\it conserved} (in fact, cyclic) current that starts and ends in the same node.
Each such a distinct integer-valued conserved current is associated with a corresponding global degeneracy resolution $({\bm k},\tilde{{\bm X}})$,  so that
\begin{eqnarray}
\label{Q-integer-component} \bar{{\bm Q}}({\bm s};{\bm k},\tilde{{\bm X}})=\sum_{a}{\bm Q}_{k_{{\bm E}(\tau_{a})}k_{{\bm E}(\tau'_{a})}}(\tilde{X}_{{\bm W}_{a}}),
\end{eqnarray}
where the summation runs over the $1$-segments for some proper partition of the time period into a set of alternating segments, and the initial and final nodes coincide. Although the definition of the integer reference current $\bar{{\bm Q}}({\bm s};{\bm k},\tilde{{\bm X}})$ involves a partition into alternating segments, we will argue in the following subsection~\ref{reference-invariant} that it does not actually depend on a particular partition choice, and therefore is fully determined by a degeneracy resolution and the driving protocol.

From Eqs.~(\ref{Q-double-limit-rational-1}) and (\ref{Q-integer-component}) we finally arrive at the averaging formula for the pumped current
\begin{eqnarray}
\label{Q-averaging} \bar{{\bm Q}}({\bm s})=\frac{1}{n_{X}}\sum_{({\bm k},\tilde{{\bm X}})}\bar{{\bm Q}}({\bm s};{\bm k},\tilde{{\bm X}})
\end{eqnarray}
that expresses $\bar{{\bm Q}}({\bm s})$ as a result of averaging the integer cyclic currents over all degeneracy resolutions with identical weights.

The fact that the integer currents in (\ref{Q-averaging}) are cyclic implies that the total pump current, $\bar{{\bm Q}}({\bm s})$, is conserved, as was expected.

\section{Winding-index formulae for quantized currents}
\label{winding-index}

In this section we present a winding index representation of the pumped current that provides an alternative way of calculating the low-temperature adiabatic limit $\bar{{\bm Q}}({\bm s})$.

\subsection{Winding index and intersection index}
\label{wind-intersect-gen}

The quantized currents in both integer and rational cases can be represented in terms of the winding numbers. As it was briefly discussed in subsection~\ref{good-bad-parmeters}, the space ${\cal M}_{X}=\widetilde{{\cal M}}_{X}\setminus D_{X}$ can be viewed as a result of withdrawing a set $D_{X}$ of bad parameters,  the latter referred to as the discriminant set. The discriminant set $D_{X}$ consists of points $({\bm E},{\bm W})\in D_{X}$ with simultaneous degeneracy of some energies $E_{{\cal J}}=E_{{\cal J}'}$ and barriers $W_{\mathfrak{A}}=W_{\mathfrak{A}'}$, and therefore has codimension ${\rm codim}(D_{X})=2$. Therefore, for a good driving protocol ${\bm s}$ that by definition avoids $D_{X}$, one can think in terms of {\it winding}, i.e. how many times the driving protocol winds around the discriminant set, such a number is usually referred to as the winding number. To formalize the notion of the winding number, one spans a surface ${\cal A}$ on ${\bm s}$, i.e, we have for its boundary $\partial{\cal A}={\bm s}$, and count the intersections of ${\cal A}$ with $D_{X}$, weighted with the proper sign factors, depending on the relative orientation of ${\cal A}$ and $D_{X}$ at the intersection point, the obtained integer number, denoted ${\cal A}\cdot D_{X}$ being known as the intersection index. The winding index is defined by ${\bm s}*D_{X}\equiv {\cal A}\cdot D_{X}$.

There are two things that should be taken care of to build a winding-index representation of the generated current. First of all, the winding index should not depend on a specific choice of the surface ${\cal A}$ spanned on ${\bm s}$. This will be achieved if $D_{X}$ is a cycle, i.e., it does not have a non-trivial boundary. Second, the winding index should be represented by a conserved current on our graph $X$ with rational components, rather than by just a single integer number. This will be achieved by extending $D_{X}$ to a cycle ${\cal D}_{X}$ with coefficients in the space of conserved currents with rational components, so that we have $\bar{{\bm Q}}({\bm s})={\bm s}\ast {\cal D}_{X}={\cal A}\cdot {\cal D}_{X}$. The cycle ${\cal D}_{X}$ is naturally referred to as the current-generating cycle.

\subsection{Geometrical structure of the discriminant and the current-generating cycle}
\label{D-X-and-calD-X}

The discriminant set $D_{X}$ has a very natural decomposition into cells, so that with each cell we can associate an {\it elementary} conserved current on our graph $X$. The set of elementary currents, associated with the cells totally determine the current-generating cycle ${\cal D}_{X}$. An elementary cell is labeled by the following data: an unordered pair $\{{\cal J},{\cal J}'\}$, related to a lowest-energy degeneracy $E_{{\cal J}}=E_{{\cal J}'}$, and an unordered pair $\{\mathfrak{A},\mathfrak{A}'\}$, related to a barrier degeneracy $W_{\mathfrak{A}}=W_{\mathfrak{A}'}$, together with an element $p\in{\cal O}({\cal X}_{1}/\{\mathfrak{A},\mathfrak{A}'\})$, i.e., an ordering $\mathfrak{B}_{1}<\ldots<\mathfrak{B}_{j}
<\{\mathfrak{A},\mathfrak{A}'\}<\mathfrak{A}_{j+2}<\ldots<\mathfrak{B}_{m}$ ``with degeneracy'' of ${\cal X}_{1}=\{\mathfrak{B}_{1},\ldots,\mathfrak{B}_{j},\mathfrak{A},\mathfrak{A}',
\mathfrak{A}_{j+2},\ldots,\mathfrak{B}_{m}\}$. An open cell $D_{{\cal J}{\cal J}',\mathfrak{A}\mathfrak{A}',p}$ consists of the parameter sets $({\bm E},{\bm W})$ that satisfy the following constraints: $E_{{\cal J}}=E_{{\cal J}}'<E_{{\cal I}}$ for any ${\cal I}\ne {\cal J},{\cal J}'$, and $W_{\mathfrak{B}_{1}}<\ldots<W_{\mathfrak{B}_{j}}<W_{\mathfrak{A}}=W_{\mathfrak{A}'}
<W_{\mathfrak{B}_{j+2}}<\ldots<W_{\mathfrak{B}_{m}}$. Stated differently in an open cell exactly two specified lowest energies are degenerate, and exactly two specified barriers are degenerate, and the ordering of all barriers is also fixed. Any open cell is contractible and topologically equivalent to an open multidimensional disc (ball). A closed cell $\bar{D}_{{\cal J}{\cal J}',\mathfrak{A}\mathfrak{A}',p}$, obtained as a topological closure of the corresponding open cell, is defined by the modified conditions, where the signs ``less'' are replaced with ``less or equal''. Naturally, different open cells do not intersect, whereas $D_{X}=\bigcup \bar{D}_{{\cal J}{\cal J}',\mathfrak{A}\mathfrak{A}',p}$. Stated differently, the discriminant $D_{X}$ can be viewed as a bunch of closed cells, glued together along their boundaries.

There is a very natural and intuitive way to associate with each cell $D_{{\cal J}{\cal J}',\mathfrak{A}\mathfrak{A}',p}$ a conserved current ${\cal D}_{{\cal J}{\cal J}',\mathfrak{A}\mathfrak{A}',p}$ by choosing a reference point $({\bm E},{\bm W})\in D_{{\cal J}{\cal J}',\mathfrak{A}\mathfrak{A}',p}$, considering a good driving protocol ${\bm s}$ that winds once around the cell staying close to the reference point, followed by evaluating the generated current and setting ${\cal D}_{{\cal J}{\cal J}',\mathfrak{A}\mathfrak{A}',p}\equiv \bar{{\bm Q}}({\bm s})$. More specifically, we choose ${\bm s}(\tau)$ by setting all parameters, except for $E_{{\cal J}}$ and $W_{\mathfrak{A}}$ being time-independent, whereas
\begin{eqnarray}
\label{s-cell} E_{{\cal J}}=E_{{\cal J}'}+\varepsilon\cos(2\pi\tau), \; W_{\mathfrak{A}}=W_{\mathfrak{A}'}+\varepsilon\sin(2\pi\tau),
\end{eqnarray}
with $\varepsilon$ being small enough not to cause any additional degeneracies. The generated current $\bar{{\bm Q}}({\bm s})$ can be readily calculated, by partitioning the time period $[-1/8,7/8]$ into four alternating $0$- and $1$-segments $[-1/8,1/8]$, $[1/8,3/8]$, $[3/8,5/8]$, and $[5/8,7/8]$. This results in
\begin{eqnarray}
\label{cal-D-components} {\cal D}_{{\cal J}{\cal J}',\mathfrak{A}\mathfrak{A}',p}&=&\frac{1}{n_{{\bm E}_{{\cal J}}}n_{{\bm E}_{{\cal J}'}}n_{{\bm W}_{\mathfrak{A}\mathfrak{A}',p}}n_{{\bm W}_{\mathfrak{A}'\mathfrak{A},p}}}\sum_{k\in{\cal J}}\sum_{k'\in{\cal J}'} \nonumber \\ &\times& \sum_{\tilde{X},\tilde{X}'}\left({\bm Q}_{kk'}(\tilde{X})+{\bm Q}_{k'k}(\tilde{X}')\right),
\end{eqnarray}
where the $\tilde{X},\tilde{X}'$ summation goes over $\tilde{X}\in\tilde{{\cal X}}_{{\bm W}_{\mathfrak{A}\mathfrak{A}',p}}$ and $\tilde{X}'\in\tilde{{\cal X}}_{{\bm W}_{\mathfrak{A}'\mathfrak{A},p}}$. The set of conserved rational-valued currents ${\cal D}_{{\cal J}{\cal J}',\mathfrak{A}\mathfrak{A}',p}$ on $X$, associated with the open cells $D_{{\cal J}{\cal J}',\mathfrak{A}\mathfrak{A}',p}$, will be referred to as the current-generating cycle, and denoted ${\cal D}_{X}$. It is convenient to represent the current-generating cycle as a formal sum
\begin{eqnarray}
\label{cal-D-formal-sum} {\cal D}_{X}=\sum_{\omega}{\cal D}_{\omega}\omega=\sum_{\omega}{\cal D}_{\omega}D_{\omega}
\end{eqnarray}
over the open cells $\omega=({\cal J}{\cal J}',\mathfrak{A}\mathfrak{A}',p)$ that decompose the discriminant set $D_{X}$.

The derived expression [Eq.~(\ref{cal-D-components})] for the elementary currents has a simple intuitive interpretation. For a given choice of $k\in {\cal J}$, $k'\in {\cal J}'$, $\tilde{X}\in\tilde{{\cal X}}_{{\bm W}_{\mathfrak{A}\mathfrak{A}',p}}$, and $\tilde{X}'\in\tilde{{\cal X}}_{{\bm W}_{\mathfrak{A}'\mathfrak{A},p}}$ we concatenate the paths $l_{kk'}(\tilde{X})$ and $l_{k'k}(\tilde{X}')$. This results in a closed path in $X$ that generates a conserved integer-valued current. Averaging these currents over all allowed choices of $(k,k',\tilde{X},\tilde{X}')$ with identical weights results in the elementary current ${\cal D}_{\omega}$, associated with the cell $\omega$.

\subsection{Pumped current as winding index}
\label{Q-as-winding}

To define the winding index ${\bm s}*{\cal D}_{X}$ with values in conserved currents on $X$ having rational components, we span the surface ${\cal A}$ on ${\bm s}$ (more specifically a disc), so that it avoids the boundaries $\bar{D}_{\omega}-D_{\omega}$ of the cells $\bar{D}_{\omega}$, i.e., the intersections are restricted to the open cells $D_{\omega}$ and are transversal, in particular, the intersection ${\cal A}\cap D_{X}$ is represented by a finite number of isolated points. Using the representation of Eq.~(\ref{cal-D-formal-sum})and the general concept of winding, briefly described in subsection~\ref{wind-intersect-gen}, we define
\begin{eqnarray}
\label{define-weighted-winding} {\bm s}*{\cal D}_{X} \equiv {\cal A}\cdot {\cal D}_{X} \equiv \sum_{\omega}{\cal D}_{\omega}\left({\cal A}\cdot D_{\omega}\right),
\end{eqnarray}
where $\omega=({\cal J}{\cal J}',\mathfrak{A}\mathfrak{A}',p)$, and the standard, i.e., integer-valued, intersection index is defined by
\begin{eqnarray}
\label{define-intersect-cell} {\cal A}\cdot D_{\omega}\equiv \sum_{{\bm x}\in {\cal A}\cap D_{\omega}}{\rm ind}_{{\bm x}}
\end{eqnarray}
and the intersection index ${\rm ind}_{{\bm x}}=\pm 1$, associated with an intersection point, reflects the relative orientation of ${\cal A}$ and $D_{\omega}$ at the intersection ${\bm x}$. Since a change in orientation of $D_{\omega}$ is accompanied by a sign change of the elementary current ${\cal D}_{\omega}$, the definition given by Eq.~(\ref{define-weighted-winding}) is invariant with respect to re-orientations of the cells $D_{\omega}$.

The winding-index formula for the low-temperature adiabatic limit of the pumped current
\begin{eqnarray}
\label{Q-winding-index} \bar{{\bm Q}}({\bm s})={\bm s}*{\cal D}_{X}
\end{eqnarray}
provides an alternative way to calculate $\bar{{\bm Q}}({\bm s})$. One can, e.g., choose some reference point ${\bm y}\in {\cal M}_{X}$ and consider ${\cal A}$ as the cone of the driving protocol ${\bm s}$ with the origin in ${\bm y}$. The obtained surface ${\bm A}$ spanned on ${\bm s}$ is smooth everywhere except for the cone origin ${\bm y}$, which is not a problem since by construction ${\bm y}$ is off the discriminant set $D_{X}$ and does not participate in the intersection. In a generic situation the intersection ${\cal A}\cap D_{X}$ is transversal, however, to convert the presented idea into an algorithm, one should introduce a specific procedure of resolving non-transverse intersections. This issue will not be addressed here.

The winding-index expression [Eq.~(\ref{Q-winding-index})] for $\bar{{\bm Q}}({\bm s})$ is very intuitive and can be rationalized as follows. Due to the robustness property of $\bar{{\bm Q}}({\bm s})$ and given a spanned surface ${\cal A}$ with a transversal intersection ${\cal A}\cap D_{X}$, the driving protocol ${\bm s}$ can be replaced by a set ${\bm s}_{{\bm x}}$ of driving protocols that wind around the intersection points ${\bm x}\in {\cal A}\cap D_{X}$, so that the pumped current remains the same, i.e., $\bar{{\bm Q}}({\bm s})=\sum_{{\bm x}\in {\cal A}\cap D_{X}}\bar{{\bm Q}}({\bm s}_{{\bm x}})$. By observing that $\bar{{\bm Q}}({\bm s}_{{\bm x}})={\cal D}_{\omega_{{\bm x}}}$, where $\omega_{{\bm x}}$ labels the cell that contains ${\bm x}\in D_{\omega_{{\bm x}}}$ and making use of Eq.~(\ref{define-weighted-winding}), we arrive at the winding index formula [Eq.~(\ref{Q-winding-index})].

\subsection{Cyclic nature of the current generating cycle and the space of robust parameters}
\label{calD-cycle-extended-M}

As stated above for the winding index ${\bm s}*{\cal D}_{X}$ to be defined properly, i.e., for the intersection index ${\cal A}\cdot {\cal D}_{X}$ to be independent of a particular choice of the surface ${\cal A}$ spanned on ${\bm s}$, the object ${\cal D}_{X}$ should be a cycle, i.e., it should not have a non-trivial boundary in some proper sense that has not been identified yet. On the other hand, the winding-index representation [Eq.~(\ref{Q-winding-index})] clearly demonstrates the invariant nature of ${\bm s}*{\cal D}_{X}$, i.e., the independence of the choice of surface ${\cal A}$, which means that referring to ${\cal D}_{X}$ as a cycle with coefficients in conserved currents on $X$ definitely makes sense.

To put the above argument on a more formal basis we note that, since our cells $D_{\omega}$ have codimension $2$, their boundaries have codimension $3$, and there are three types of boundaries: (i) triple energy degeneracy together with a simple barrier degeneracy, i.e., $E_{{\cal J}}=E_{{\cal J}'}=E_{{\cal J}''}$ and $W_{\mathfrak{A}}=W_{\mathfrak{A}'}$, (ii) simple energy degeneracy together with a triple barrier degeneracy, i.e., $E_{{\cal J}}=E_{{\cal J}'}$ and $W_{\mathfrak{A}}=W_{\mathfrak{A}'}=W_{\mathfrak{A}''}$, and (iii) simple energy degeneracy together with a pair of distinct barrier degeneracies, i.e., $E_{{\cal J}}=E_{{\cal J}'}$, $W_{\mathfrak{A}}=W_{\mathfrak{A}'}$, and $W_{\mathfrak{B}}=W_{\mathfrak{B}'}$. Of course, we need to specify the orderings of the degenerate groups of barriers together with the non-degenerate barriers, which represents the total boundary as a union of closed cells of codimension $3$, whose intersections correspond to higher-order degeneracies, having codimension $4$, and will not play role in our consideration. A type-(i) boundary cell has $6$ current-carrying cells attached to it, e.g., $E_{{\cal J}}=E_{{\cal J}'}<E_{{\cal J}''}$ and $W_{\mathfrak{A}}=W_{\mathfrak{A}'}$, with the ordering with respect to the other barriers being determined by their ordering in the considered boundary cell. The situation can be illustrated in a $3$-dimensional space of {\it relevant} parameters $(E_{{\cal J}'}-E_{{\cal J}})$, $(E_{{\cal J}''}-E_{{\cal J}})$, and $(W_{\mathfrak{A}'}-W_{\mathfrak{A}})$. Similarly, a type-(ii) boundary cell has $6$ current-carrying cells attached to it. The relevant parameters are $(E_{{\cal J}'}-E_{{\cal J}})$, $(W_{\mathfrak{A}'}-W_{\mathfrak{A}})$, and $(W_{\mathfrak{A}''}-W_{\mathfrak{A}})$. A type-(iii) boundary cell shares the boundary of $4$ cells, e.g., $E_{{\cal J}}=E_{{\cal J}'}$, $W_{\mathfrak{A}}<W_{\mathfrak{A}'}$, and $W_{\mathfrak{B}}=W_{\mathfrak{B}'}$, with the relevant parameters $(E_{{\cal J}'}-E_{{\cal J}})$, $(W_{\mathfrak{A}'}-W_{\mathfrak{A}})$, and $(W_{\mathfrak{B}'}-W_{\mathfrak{B}})$. 

The independence of the weighted winding index ${\bm s}*{\cal D}_{X}={\cal A}\cdot {\cal D}_{X}$ on a particular choice of the spanned surface, which actually takes place due to Eq.~(\ref{Q-winding-index}), boils down to the condition that for each boundary cell, the sum of elementary currents ${\cal D}_{\omega}$, associated with the cells $D_{\omega}$ attached to it, should be equal to zero. Of course, all ${\cal D}_{\omega}$ should be accounted for with the proper signs that, e.g., correspond to all elementary currents to be incoming to the boundary. Stated differently we can introduce the boundary operator that associates with ${\cal D_{X}}=\sum_{\omega}{\cal D}_{\omega}$ a formal sum over the boundary cells $B_{\varrho}$, labeled by $\varrho$:
\begin{eqnarray}
\label{define-boundary-D-X} \partial{\cal D}_{X}=\partial\sum_{\omega}{\cal D}_{\omega}\omega=\sum_{\omega}{\cal D}_{\omega}\partial\omega=\sum_{\alpha}\varrho\sum_{\omega\supset\varrho}{\cal D}_{\omega},
\end{eqnarray}
where $\omega\supset\varrho$ naturally means $\bar{D}_{\omega}\supset {\cal B}_{\varrho}$, and Eq.~(\ref{define-boundary-D-X}) is consistent with a natural definition $\partial\omega=\sum_{\varrho\subset\omega}\varrho$. Using this notation we can reformulate the requirement of the winding index ${\bm s}*{\cal D}_{X}$ to be well-defined in two equivalent forms (invariant and component)
\begin{eqnarray}
\label{cycle-condition} \partial{\cal D}_{X}=0, \;\;\; \sum_{\omega\supset\varrho}{\cal D}_{\omega}=0,
\end{eqnarray}
with the invariant form claiming that the boundary of ${\cal D}_{X}$ is trivial, i.e., ${\cal D}_{X}$ is a cycle, and the component form to be some kind of a higher dimensional Kirchhoff law, claiming that the sum of currents, represented by ${\cal D}_{\omega}$ that are incoming to any boundary component ${\cal B}_{\varrho}$ is equal to zero.

The derived winding index representation allows the subspace ${\cal M}_{X}$ of good parameters to be readily extended to a space ${\cal M}_{X}^{r}\supset {\cal M}_{X}$, described as the subset of robust parameters ${\cal M}_{X}^{r}\subset \tilde{{\cal M}}_{X}$, i.e, that provides robust behavior of the low-temperature adiabatic current. Noting that only those cells $D_{\omega}$ with ${\cal D}_{\omega}\ne 0$ participate in the current generation, we have:
\begin{eqnarray}
\label{extend-M-rational} {\cal M}_{X}^{r}={\cal M}_{X}-D_{X}^{r}, \;\;\; D_{X}^{r}\equiv \bigcup_{\omega}^{{\cal D}_{\omega}\ne 0}\bar{D}_{\omega}.
\end{eqnarray}

\subsection{Topological view and invariant nature of the reference currents}
\label{reference-invariant}

In this subsection we present some simple and intuitive topological arguments that support the assertion that a reference current $\bar{{\bm Q}}({\bm s};{\bm k},\tilde{{\bm X}})$ 
is independent of the particular choice of the partition of the protocol period into alternating segments. We will also apply them to demonstrate that the pumped current $\bar{{\bm Q}}({\bm s})$ is {\it robust}, i.e., it does not change upon continuous deformations of the driving protocol ${\bm s}$, as long as ${\bm s}$ entirely stays in the subspace ${\cal M}_{X}$ of good parameters. Our arguments are based on introducing an alternative invariant definition of the reference currents $\bar{{\bm Q}}({\bm s};{\bm k},\tilde{{\bm X}})$ that does not involve any partitions.

To that end consider the space ${\cal M}_{X}\times |X|$, where $|X|$ denotes the geometrical realization of our graph $X$, i.e., the nodes of $X$ are represented by points in $|X|$, whereas the links are represented by segments, whose ends are attached to the points that represent the boundary nodes. In simple words $|X|$ is a one-dimensional space, obtained when we ``draw'' a graph. We can consider a stochastic trajectory of a particle as a trajectory of a point $({\bm x},{\bm\eta})\in {\cal M}_{X}\times |X|$, where the component ${\bm x}$ moves in the ``good'' parameter space ${\cal M}_{X}$ deterministically according to the driving protocol, whereas ${\bm\eta}$ moves in a stochastic manner in the graph space $|X|$. We fix a degeneracy resolution $({\bm k},\tilde{{\bm X}})$ and further impose two restrictions on the particle motion: (i) when ${\bm x}=({\bm E},{\bm W})\in U_{1}$, i.e., the barrier data is non-degenerate, the particle position ${\bm\eta}$ on the graph is restricted to ${\bm\eta}\in |\tilde{X}_{{\bm W}}|\subset |X|$, (ii) when ${\bm x}$ does not belong to $U_{1}$, i.e., the barrier data degeneracy occurs, and hence ${\bm x}\in U_{0}$, the particle should be located at the node $k_{{\bm E}}$, i.e. ${\bm\eta}=k_{{\bm E}}$.

Consider a periodic stochastic trajectory $({\bm s}(\tau),{\bm\eta}(\tau))$ in the restricted space, where ${\bm s}(\tau)$ is the driving protocol over a time period. The ${\bm\eta}(\tau)$ component of the trajectory can be arbitrary, provide the above restrictions are satisfied. Despite a large variety of allowed trajectories, the currents generated by all these trajectories are identical. There is a topological reason for that: the particle motion along the graph $X$ is always restricted to a tree: ${\bm\eta}(\tau)\in |\tilde{X}_{{\bm W}(\tau)}|$ when ${\bm s}(\tau)\in U_{1}$, and ${\bm\eta}(\tau)=k_{{\bm E}}(\tau)$, otherwise. Since trees are contractible, all trajectories ${\bm\eta}(\tau)$ for the same driving protocol are topologically (homotopically) equivalent, and hence produce the same current, denoted $\bar{{\bm Q}}({\bm s};{\bm k},\tilde{{\bm X}})$. To demonstrate that the current $\bar{{\bm Q}}({\bm s};{\bm k},\tilde{{\bm X}})$, defined above in an invariant manner using simple topological terms, is identical to the one defined by Eq.~(\ref{Q-integer-component}) in terms of a partitioning, we evaluate the ``topological'' current for a special choice of the trajectory ${\bm\eta}$, so that $({\bm s},{\bm\eta})$ goes along the restricted subspace. For a $0$-segment we set ${\bm\eta}(\tau)=k_{{\bm E}}(\tau)$, for a $1$-segment $I_{a}$ the trajectory ${\bm\eta}(\tau)$ connects $k_{{\bm E}}(\tau_{a})$ to $k_{{\bm E}}(\tau'_{a})$ via the path $l_{k_{{\bm E}}(\tau_{a})k_{{\bm E}}(\tau'_{a})}(\tilde{X}_{{\bm W}_{a}})$. By inspecting the definitions we see that the ``topological'' current generated by the described trajectory ${\bm\eta}$ reproduces Eq.~(\ref{Q-integer-component}).

The topological view of the reference currents $\bar{{\bm Q}}({\bm s};{\bm k},\tilde{{\bm X}})$ combined with the averaging formula allows the robustness property to be demonstrated in a simple and intuitive way. Indeed, according to the above a reference current $\bar{{\bm Q}}({\bm s};{\bm k},\tilde{{\bm X}})$ can be obtained by lifting the closed path ${\bm s}$ to the restricted subspace of ${\cal M}_{X}\times |X|$ (the lift being unique up to a homotopy), followed by projecting the lift into $|X|$ with evaluating the current,
associated with the obtained path in the graph $X$. All operations preserve homotopy: if ${\bm s}$ and ${\bm s}'$ are {\it homotopic}, i.e., ${\bm s}$ can be deformed to ${\bm s}'$ in a continuous manner, then so are their lifts, and further the projections of the lifts into $|X|$. Since the conserved current associated with a closed path in the graph $X$ depends on the homotopy equivalence class of the path only, any reference current is a homotopy invariant of the driving protocol ${\bm s}$, i.e., is robust. This implies the robustness of $\bar{{\bm Q}}({\bm s})$ via the averaging formula [Eq.~(\ref{Q-averaging})].

\subsection{Winding index representation and current tubes}
\label{winding-index-tubes}

Spanning a surface ${\cal A}$ on the driving protocol ${\bm s}$ and applying the Stokes theorem, we can express the current ${\bm Q}({\bm s};\beta)$ generated at finite temperature
\begin{eqnarray}
\label{J-adiabatic-surface-Stokes} {\bm Q}({\bm s};\beta)&=&\int_{{\cal A}}d{\bm
A}\wedge d{\bm\rho}^{B}=\int_{{\cal A}}{\bm F} \nonumber \\ &=&\sum_{j\alpha}\int_{{\cal A}}{\bm F}^{\alpha j}({\bm E},{\bm W};\beta)dE_{j}\wedge dW_{\alpha},
\end{eqnarray}
as a surface integral of the (antisymmetric) curvature tensor ${\bm F}$ that takes values in the vector space of conserved currents on $X$ and has the form
\begin{eqnarray}
\label{F-adiabatic} {\bm F}^{\alpha j}({\bm E},{\bm W};\beta)=\sum_{k\in X_{0}}\frac{\partial{\bm A}^{k}({\bm W};\beta)}{\partial W_{\alpha}}
\frac{\partial{\rho_{k}^{B}({\bm E};\beta)}}{\partial E_j}.
\end{eqnarray}

The representation of Eq.~(\ref{J-adiabatic-surface-Stokes}) expresses the adiabatic (and not necessarily low-temperature) current as a flux of ${\bm F}$ through a spanned surface ${\cal A}$, and therefore provides a physical interpretation of the winding index representation of the low-temperature limit. Indeed, Eq.~(\ref{Q-winding-index}) with the winding index defined by Eq.~(\ref{define-weighted-winding}) as the weighted intersection index ${\cal A}\cdot{{\cal D}}_{X}$, the low-temperature current can be viewed as the flux of some curvature ${\bm F}$, which is restricted to the discriminant set $D_{X}^{r}$, and hence is described by a $\delta$-function type distribution. This is not surprising, since it describes the low-temperature $\beta\to\infty$ behavior of the curvature tensor ${\bm F}$, whose explicit expression for arbitrary $\beta$ is given by Eq.~(\ref{F-adiabatic}). For low temperature, ${\bm F}$ is substantially concentrated in a narrow tube around $D_{X}^{r}$, with the tube transverse size tending to zero exponentially with $\beta\to\infty$, whereas the flux through a small transverse surface that intersects $D_{X}^{r}$ at a point ${\bm x}\in D_{\omega}$ reaching a finite limiting value equal to ${\cal D}_{\omega}$. This can be readily established by  comparing Eq.~(\ref{F-adiabatic}) with the the explicit expressions [Eqs.~(\ref{q-explicit-adiabatic}) and (\ref{gamma-distr-adiabatic})], provided by the Kirchhoff theorem, which results in
\begin{eqnarray}
\label{F-adiabatic-2} {\bm F}^{\alpha j}({\bm E},{\bm W};\beta)=\sum_{k,\tilde{X}}{\bm Q}^{k\tilde{X}}\frac{\partial{\bm\varrho}_{\tilde{X}}^{B}({\bm W};\beta)}{\partial W_{\alpha}}
\frac{\partial{\rho_{k}^{B}({\bm E};\beta)}}{\partial E_j}.
\end{eqnarray}
followed by a straightforward study of the low-temperature behavior of the weighted curvature ${\bm F}$.

Naturally the cyclic property $\partial{\cal D}_{X}=0$ of the current generating cycle ${\cal D}_{X}$ is nothing else than the low-temperature counterpart of corresponding property $d{\bm F}(\beta)=0$, the latter holding for all temperatures.

\section{Discussion}
\label{conclusion}

We have considered the current generated in a stochastic network due to periodic time-dependence of the kinetic rates. We demonstrated that for a typical driving protocol its low-temperature adiabatic limit $\bar{{\bm Q}}({\bm s})$ is well-defined and robust, i.e., does not change at all with respect to moderate perturbations of the driving protocol. We established that the integer quantization of this current follows from the fact that 
in the considered low temperature limit no average current flows over the largest barrier if the corresponding link can be removed without breaking graph into disjoined component. This property 
holds true even if an alternative path forces the system to overcome a series of barriers and even though adiabatic limit assumes that system performs many random transitions through all graph links. As a result, effectively, motion
on the graph is restricted to a set of spanning trees with identical weights. 
Using this observation, we derived an explicit expression for the value of the fractional quantized current in a general graph and for a general protocol. 

Our main result - the averaging formula, given by Eq.~(\ref{Q-averaging}), provides a combinatorial approach to calculation of the fractionally quantized currents. Interestingly, it tells that fractional quantization is reduced to integer quantization case, namely, the final result corresponds to averaging over all properly defined legitimate degeneracy resolutions of the barriers and node energies, with the individual contributions, labeled by the degeneracy resolutions being represented by integer quantized cyclic currents that can be found using a deterministic picture for integer current calculation. This simplification can be used, e.g., for efficient numerical calculations of fractionally quantized currents in large networks, avoiding time-expensive matrix inversion and diagonalization that are used in pump current calculations \cite{sinitsyn-07epl,jarzynski-08prl,ohkubo-08jstat}.

We have also established the winding index view of the current quantization in the low-temperature adiabatic limit. This is achieved by identifying the cellular structure of the space $D_{X}^{r}\subset \widetilde{M}_{X}$ of non-robust parameters, as well as associating with each cell of maximal dimension (which corresponds to codimension $2$) a conserved current, so that for each cell of codimension $3$ the sum of the currents (with a proper sign choice), associated with the codimension $2$ cells attached to it, is zero. The above identification leads to a notion of the current generating cycle ${\cal D}_{X}$ with the coefficients in the conserved current, so that the current $\bar{{\bm Q}}({\bm s})$ generated by a robust driving protocol is given by the winding index of ${\bm s}$ and ${\cal D}_{X}$, as given by Eq.~(\ref{Q-winding-index}). The winding index representation is reminiscent of the Aharonov-Bohm effect in quantum mechanics, or strictly speaking its generalization to the case of arbitrary dimension of the space $\widetilde{{\cal M}}_{X}$, rather than just $3$. In the latter case the space $D_{X}^{r}$ is represented by a graph, with the conserving currents residing on the links and the current conservation at the nodes (the sum of currents entering any node is equal to zero). We showed how to determine the space $D_{X}^{r}$ explicitly, and provided a procedure that assigns a quantized conserving current with each cell of $D_{X}^{r}$, resulting in the cycle ${\cal D}_{X}$. Our work provides a direct algorithmic approach for a complete classification of fractionally quantized currents in any specific model of stochastic Markovian motion of a mesoscopic system under an  influence of a periodic driving protocol.

We conclude with noting that in section~\ref{fractional-Q} we implemented the Kirchhoff theorem to develop a useful tool for applying the low-temperature adiabatic limit, i.e., we derived an explicit expression [Eqs.~(\ref{q-geom2-explicit}) and (\ref{define-gamma-distr})] for the average generated current for a general situation, i.e., finite values of the inverse temperature $\beta$ and driving protocol period $\tau_{D}$. This expression might have its own value, especially in the adiabatic $\tau_{D}\to \infty$, still finite-temperature case, where the expressions [Eqs.~(\ref{q-explicit-adiabatic}) and (\ref{gamma-distr-adiabatic})] show some surprising symmetry between the sets of nodes and spanning trees of our original network $X$. This symmetry could be a manifestation of some hidden duality, with the nodes of the dual graph defined as the spanning trees of the original one. We note that on cyclic graphs
such a duality is well known \cite{Solich}. In particular, it was used to derive unusual fluctuation relations for currents  \cite{sinitsyn-ren-11jstat}.
 The existence of such duality in models defined on generic graphs remains an open question and could be subject of future work.


\begin{acknowledgments}
{\it  This
material is based upon work supported by NSF under Grants No. CHE-0808910 and ECCS-0925618. N.A.S. was additionally supported by DOE under
Contract No.\ DE-AC52-06NA25396.}
\end{acknowledgments}

\appendix

\section{Properties of minimal spanning trees}
\label{min-span-details}

In this appendix, we rationalize the properties of the minimal spanning trees, described in subsection~\ref{minimal-spanning} for the integer quantization setting, as well as in subsection~\ref{min-span-structures} for rational quantization setting. More specifically we are dealing with property (i) in the integer case, and property (ii) for the rational quantization setting. We reiterate that the difference between the two cases is that in the integer case the minimal spanning tree is unique, and is denoted by $\tilde{X}_{{\bm W}}$, whereas in the rational case there is a set of minimal spanning trees, denoted by $\tilde{{\cal X}}_{{\bm W}}$ with all minimal spanning trees having the same value of the energy function $W(\tilde{X})$. The property under consideration can be formulated in a unique fashion for both integer (non-degenerate) and rational (degenerate) cases: The value of the energy function of a minimal spanning tree is strictly lower than the energy value of any spanning tree that is not minimal. Although the non-degenerate (integer) situation is a particular example of the degenerate (rational) counterpart, for the sake of presentation clarity we start with the non-degenerate situation, followed by demonstrating how the presented arguments can be extended to the degenerate case. Note that in both cases it is sufficient to show that if a spanning tree $\tilde{X}$ is not a minimal spanning tree, then there is a spanning tree $\tilde{X}'$ with $W(\tilde{X}')< W(\tilde{X})$.

We start with the non-degenerate case. Let $(\alpha_{1},\ldots,\alpha_{k})$ and $(\beta_{1},\ldots,\beta_{k})$ be the elements of $X_{1}\setminus (\tilde{X}_{{\bm W}})_{1}$ and $X_{1}\setminus \tilde{X}_{1}$, i.e., the links withdrawn to obtain the trees $\tilde{X}_{{\bm W}}$ and $\tilde{X}$, respectively, in the decreasing order of their barrier levels. The latter means $W_{\alpha_{1}}> \ldots> W_{\alpha_{k}}$ and $W_{\beta_{1}}> \ldots> W_{\beta_{k}}$. Let $j$ be the smallest index such that $\alpha_{j}\ne \beta_{j}$, and $\alpha_{i}=\beta_{i}$ for $i< j$. Consider a spanning subgraph $X'\subset X$, obtained from $X$ by withdrawing the edges $\alpha_{1},\ldots,\alpha_{j-1}$, or equivalently $\beta_{1},\ldots,\beta_{j-1}$. By the definition of $\tilde{X}_{{\bm W}}$ the edge $\alpha_{j}$ is not a bridge of $X'$ (i.e., its withdrawal does not disconnect the graph), and $W_{\beta_{i}}< W_{\alpha_{j}}$ for $i\ge j$. Let $\tilde{X}^{1}$ and $\tilde{X}^{2}$ be the two trees obtained from $\tilde{X}$ by withdrawing the edge $\alpha_{j}$. Then there is at least one edge, say $\beta_{s}$ among $\beta_{j},\ldots,\beta_{k}$ that connects $\tilde{X}^{1}$ to $\tilde{X}^{2}$, since otherwise the edge $\alpha_{j}$ would be a bridge of $X'$. Therefore, replacing the edge $\beta_{s}$ with $\alpha_{j}$ in $\tilde{X}$ results in another spanning tree, denoted $\tilde{X}'$ with $W(\tilde{X}')< W(\tilde{X})$, since $W_{\alpha_{j}}> W_{\beta_{s}}$.

The above arguments may be extended to the degenerate case in a straightforward way. To that end, we note that since the number of links in each group $\mathfrak{A}\subset X_{1}$ of degenerate barriers that are withdrawn to obtain a minimal spanning tree is the same for all minimal spanning trees, we can consider the groups that provided at least one link to the withdrawal procedure and put them in the decreasing order $(\mathfrak{A}_{1},\ldots,\mathfrak{A}_{k})$ of barrier values $W_{\mathfrak{A}_{1}}> \ldots > W_{\mathfrak{A}_{k}}$. We also denote by $(n_{1},\ldots,n_{k})$ the numbers of links withdrawn in each group to obtain a minimal spanning tree (as noted above, these numbers are the same for all minimal spanning trees). We further consider a spanning tree $\tilde{X}$ that is not minimal (i.e., does not belong to $\tilde{{\cal X}}_{{\bm W}}$) and denote by $(\mathfrak{B}_{1},\ldots,\mathfrak{B}_{l})$ and $(m_{1},\ldots,m_{l})$ the groups (in the decreasing order of the barrier levels) and the numbers of links, respectively, that provided to the withdrawing procedure that resulted in $\tilde{X}$. Let $j$ be the smallest number so that $(\mathfrak{A}_{j},n_{j})\ne (\mathfrak{B}_{j},m_{j})$, and $(\mathfrak{A}_{i},n_{i})= (\mathfrak{B}_{i},m_{i})$ for $i< j$. Let $X'\subset X$ be a spanning subgraph, obtained from $X$ by withdrawing all edges that belong to $\bigsqcup_{i=1}^{j}\mathfrak{A}_{i}\setminus \tilde{X}_{1}$. Since, by the construction of minimal spanning trees, ${\rm card}(\mathfrak{A}_{j}\setminus \tilde{X}_{1})< n_{j}$, there is always a link $\alpha\in \mathfrak{A}_{j}$, which does not belong to $X'_{1}\setminus \tilde{X}_{1}$, and therefore $W_{\alpha}> W_{\beta}$ for all $\beta\in X'_{1}\setminus \tilde{X}_{1}$. Similar to the integer case we can replace the link $\alpha$ that belongs to $\tilde{X}_{1}$ with some properly chosen link $\beta$ from $X'_{1}\setminus \tilde{X}_{1}$ to obtain a tree $\tilde{X}'$ with $W(\tilde{X}')< W(\tilde{X})$.

%

\newpage

\appendix


\begin{thebibliography}{99}


\bibitem{westerhoff-86} H.\ V.\ Westerhoff et al., {\em
    Proc. Natl. Acad. Sci.\ U.S.A.} {\bf 83}, 4734 (1986);
P. H\"anggi, F. Marchesoni, {\it Rev. Mod. Phys.} {\bf 81}, 387 (2009);
 R.\ D.\ Astumian et al.,  {\em
   Phys. Rev. A} {\bf 39}, 6416 (1988).
 J.~M.~R.~Parrondo, {\em Phys.\ Rev.\ E} {\bf 57},
  7297 (1997); R.\ D.\ Astumian and I.\ Derenyi, {\em Eur.\
    Biophys.\ J.} {\bf 27}, 474 (1998); R.\ D.\ Astumian and I.\ Derenyi, {\em Eur.\
    Biophys.\ J.} {\bf 27}, 474 (1998); Y. V. Pershin {\it et al}. {\it Appl. Phys. Lett.} {\bf 95} 022114 (2009); R.\ D.\ Astumian and P.\ H\"anggi, {\em Phys.\
    Today} {\bf 55}, 33 (2002); Y. Otwinowski, S. Tanase-Nicola, and I. Nemenman, J. Stat. Phys, {\bf 144}, 367 (2011).



\bibitem{brouwer-98} P.\ W.\ Brouwer, {\em Phys. Rev. B} {\bf 58}, R10135 (1998).

\bibitem{makhlin-mirlin-01} Y.\ Makhlin and A.\ D.\ Mirlin, { Phys. Rev. Lett.} {\bf 87}, 276803 (2001).


\bibitem{moskalets-buttiker-02} M.\ Moskalets  and M.\ B\"{u}ttiker, { Phys. Rev. B} {\bf 66}, 035306 (2002).

\bibitem{pump_berry} J.\ E.\ Avron, {\em Phys. Rev. B} {\bf 62}, R10618 (2000); Q.\ Niu and D.\ J.\ Thouless, { J. Phs. A: Math. Gen.} {\bf 17}, 2453 (1984).


\bibitem{kamenev} A. Andreev and A. Kamenev
{ Phys. Rev. Lett. } {\bf 85}, 1294 (2000).

\bibitem{levitov-02} L. S. Levitov, "Quantum Noise in Mesoscopic Systems," ed. Yu V Nazarov (Kluwer, 2003),  Preprint: arXiv/0210284 (2002).

\bibitem{turnstile} L. G. Geerligs {\it et al.}, {Phys. Rev. Lett.} {\bf 64}, 2691 (1990).

\bibitem{buttiker-1} M. Albert, C. Flindt, and M. B\"uttiker
Phys. Rev. B {\bf 82}, 041407 (2010); M. Albert, C. Flindt, and M. B\"uttiker, Invited
  contribution to ICNF 2011, Toronto, Canada, June, 2011,
  arXiv:1102.2343 (2001).

\bibitem{shi} Y. Shi and Q. Niu,  Europhys. Lett. {\bf 59},  324
  (2002).



\bibitem{astumian-07pnas}  D. Astumian, { Proceed. Nat. Acad. Sci. U.S.A.} {\bf 104}, 19715 (2007).

\bibitem{Leigh-03} D. A. Leigh {\it et al.}, { Nature (London)}
  {\bf 424}, 174 (2003).

\bibitem{astumian-11rev} D. R. Astumian, Annu. Rev. Biophys. {\bf 40} 289–313 (2011).



\bibitem{sinitsyn-09review}
N.~A. Sinitsyn, { J. Phys. A: Theor. Comp.} {\bf 42}, 193001 (2009).



\bibitem{sinitsyn-09jcp} V. Y. Chernyak and N. A. Sinitsyn, J. Chem. Phys. {\bf 131},
181101 (2009).

\bibitem{sinitsyn-11jcp} A. V. Akimov and N. A. Sinitsyn, J. Chem. Phys. {\bf 135}, 224104 (2011).

\bibitem{hanggi-10}  J. Ren, P. H\"anggi, and B. Li,
  Phys. Rev. Lett. {\bf 104}, 170601 (2010); J. Ren and B. Li,
  Phys. Rev. E {\bf 81}, 021111 (2010).



\bibitem{sinitsyn-11jpa} S. Rahav, J. Stat. Mech.  P09020 (2011); K. Sekimoto, F. Takagi and T. Hondou, Phys. Rev. E {\bf 62}, 7759 (2000); B. Gaveau, M. Moreau and L. S. Schulman  Phys. Rev. Lett. {\bf 105} 060601 (2010); M. Santillan, H. Qian, Phys. Rev. E {\bf 83}, 041130 (2011); J. Prost, J-F. Joanny and J. M. R. Parrondo,  Phys. Rev. Lett. {\bf 103} 090601 (2009); N A Sinitsyn, J. Phys. A: Math. Theor. {\bf 44}, 405001 (2011); M. Polettini, EPL, arXiv/1110.0608 (2011); T. Sagawa, and H. Hayakawa, Phys. Rev. E {\bf 84}, 051110 (2011); K. Saito, H. Tasaki, arXiv/1105.2168 (2011); E. Boksenbojm, C. Maes, K. Netocny, and J Pesek, EPL {\bf 96}  40001 (2011); J. Pesek, E. Boksenbojm, K. Netocny, Preprint: arXiv/1111.5566; K. Hovhannisyan and A. E Allahverdyan, J. Stat. Mech.  P06010 (2010).


\bibitem{cao} J. Cao, JPC B, {\bf 110}, 19040 (2006).


\bibitem{FCS-QI} I. Klich and L. Levitov, Phys. Rev. Lett. {\bf 102}, 100502 (2009). 



\bibitem{phase-tr-FCS} J. P. Garrahan and I. Lesanovsky, , Phys. Rev. Lett. {\bf 104}, 160601 (2010). 

\bibitem{abanov-10epl} D. A. Ivanov, and A. G. Abanov, EPL {\bf 92}, 37008 (2010).

\bibitem{sinitsyn-10jstat} V. Y. Chernyak, N. A. Sinitsyn,  J. Stat. Mech. L07001, (2010).

\bibitem{sinitsyn-11pre} N. A. Sinitsyn, A. Akimov, V. Y. Chernyak,
 { Phys. Rev.} E {\bf 83}, 021107 (2011).

 \bibitem{sinitsyn-ren-11jstat}  J. Ren, V. Y. Chernyak and N. A. Sinitsyn, {J. Stat. Mech}  P05011 (2011).

\bibitem{jarzynski-11jstat} D. Mandal and C. Jarzynski, JSTAT P10006 (2011).



\bibitem{netocny-10} C. Maes, K. Netocny, S. R. Thomas,  J. Chem. Phys. {\bf 132}, 234116 (2010).

\bibitem{horowitz-09} J. E. Horowitz and C. Jarzynski {
    J. Stat. Phys.} {\bf 136}, 917 (2009).


\bibitem{jarzynski-08prl} S. Rahav, J. Horowitz and C. Jarzynski, { Phys. Rev. Lett.} {\bf 101}, 140602 (2008).

\bibitem{sinitsyn-08prl}  V. Y. Chernyak and N. A. Sinitsyn {
    Phys. Rev. Lett.} {\bf 101} 160601 (2008).



\bibitem{pistolesi} F. Pistolesi
{\it Phys. Rev.} B {\bf 69}, 245409 (2004).

\bibitem{pump-class} D. Meidan, T. Micklitz, P. W. Brouwer, Preprint: arXiv/1107.2215 (2011).

\bibitem{hill-book} T. L. Hill, ``Free energy Transduction and
  Biochemical Cycle Kinetics'', Dover Publ., INC. Mineola, New York
  (2004).

\bibitem{graph-book} B. Bollobas, ``Modern Graph Theory'',
  Springer-Verlag, New York, Berlin, Heidelberg (1998).

\bibitem{sinitsyn-07epl} N. A. Sinitsyn and I. Nemenman { Euro. Phys. Lett.} {\bf 77} 58001 (2007).
\bibitem{ohkubo-08jstat}  J. Ohkubo,  { J. Stat. Mech.} P02011 (2008).
\bibitem{Solich} R. L. Jack and P. Sollich,  { J. Stat. Mech.} P11011 (2009).

\end{thebibliography}
\end{document}